\documentclass[journal=tosc]{iacrtrans}

\usepackage{cite}
\usepackage{amsmath,amssymb,amsfonts}
\usepackage{graphicx}
\usepackage{textcomp}
\usepackage[dvipsnames]{xcolor}
\usepackage[hyphens]{url}
\usepackage{fancyhdr}
\usepackage{hyperref}

\usepackage{comment}
\usepackage{booktabs}
\usepackage{subfig}
\usepackage{amsmath}
\usepackage{array} 
\newcolumntype{P}[1]{>{\centering\arraybackslash}p{#1}}
\newcolumntype{M}[1]{>{\centering\arraybackslash}m{#1}}
\usepackage{caption}
\usepackage{wrapfig}
\usepackage{algpseudocode}
\algrenewcommand{\algorithmiccomment}[1]{\hfill\textbf{//}\,#1}
\usepackage{algorithm}
\algrenewcommand\alglinenumber[1]{\small #1:}
\usepackage{circledsteps}
\pgfkeys{/csteps/inner color=white}
\pgfkeys{/csteps/fill color=black}
\usepackage{pifont}
\newcommand{\cmark}{\ding{51}}%
\newcommand{\xmark}{\ding{55}}%
\usepackage{multirow}
\usepackage{mdframed}
\usepackage{makecell}
\usepackage{listings}
\usepackage{verbatim}

\lstdefinelanguage{LLVM}{
    keywords={call, ptr, i32, void},
    morecomment=[l]{;},
    sensitive=true
}

\lstdefinelanguage{CNC-Assembly}{
    keywords={[1]ld_cmd, rd_d2cnc, sw_cnc, aes_cnc, keccak_cnc, ntt_cnc, kyber_cnc, dilithium_cnc},
    keywords={[2]li, la, lw, sw, add, sub, jr, jalr},
    keywords={[3]t0, t1, t2, t3, t4, t5, t6, a0, a1, a2, a3, s0, s1, sp, ra},
    morecomment=[l]{\#},
    keywordstyle={[1]\bfseries\color{BrickRed}}, 
    keywordstyle={[2]\bfseries\color{NavyBlue}}, 
    keywordstyle={[3]\color{purple}},            
    commentstyle=\itshape\color{OliveGreen},
    identifierstyle=\color{black},                
    sensitive=false,
    morestring=[b]"
}

\usepackage{soul}
\newcommand{\hlc}[2][yellow]{{%
    \colorlet{foo}{#1}%
    \sethlcolor{foo}\hl{#2}}%
}

\definecolor{r1color}{RGB}{0, 0, 255}      
\definecolor{r2color}{RGB}{0, 128, 0}      
\definecolor{r3color}{RGB}{128, 0, 128}    
\definecolor{r4color}{RGB}{255, 140, 0}    
\definecolor{addcolor}{RGB}{139, 69, 19}   

\newcommand{\revised}[1]{#1}

\newcommand{\revisedrone}[1]{#1}
\newcommand{\revisedrtwo}[1]{#1}
\newcommand{\revisedrthree}[1]{#1}
\newcommand{\revisedrfour}[1]{#1}
\newcommand{\revisedadd}[1]{#1}
\newcommand{\added}[1]{#1}
\newcommand{\revisionnote}[1]{}
\newcommand{\revisionnoter}[2]{}
\newcommand{\revisionnoteadd}[1]{}

\usepackage{pdfcomment}

\edef\oldtt{\ttdefault}
\usepackage[scaled]{beramono}
\usepackage[T1]{fontenc}
\renewcommand{\ttdefault}{\oldtt}
\newcommand{\bera}[1]{\text{\fontfamily{fvm}\selectfont #1}}
\usepackage{mathtools}

\usepackage{setspace}
\usepackage{tikz}
\usepackage{pifont}
\usepackage[most]{tcolorbox}

\author{Jingyao Zhang\inst{1}, Elaheh Sadredini\inst{2}}
\institute{
  University of California, Riverside, \email{jzhan502@ucr.edu}
  \and
  University of California, Riverside, \email{elahehs@ucr.edu}
}

\title{A Near-Cache Architectural Framework for Cryptographic Computing}

\begin{document}


\maketitle


\begin{abstract}
Recent advancements in post-quantum cryptographic algorithms have led to their standardization by the National Institute of Standards and Technology (NIST) to safeguard information security in the post-quantum era. These algorithms, however, employ public keys and signatures that are 3 to 9$\times$ longer than those used in pre-quantum cryptography, resulting in significant performance and energy efficiency overheads. A critical bottleneck identified in our analysis is the cache bandwidth. 
This limitation motivates the adoption of on-chip in-/near-cache computing, a computing paradigm that offers high-performance, exceptional energy efficiency, and flexibility to accelerate post-quantum cryptographic algorithms.
Our analysis of existing works reveals challenges in integrating in-/near-cache computing into modern computer systems and performance limitations due to external bandwidth limitation, highlighting the need for innovative solutions that can seamlessly integrate into existing systems without performance and energy efficiency issues.
In this paper, we introduce a near-cache-slice computing paradigm with support of customization and virtual address, named Crypto-Near-Cache (CNC), designed to accelerate post-quantum cryptographic algorithms and other applications. 
By placing SRAM arrays with bitline computing capability near cache slices, high internal bandwidth and short data movement are achieved with native support of virtual addressing.
An ISA extension to facilitate CNC is also proposed, with detailed discussion on the implementation aspects of the core/cache datapath. 

\end{abstract}

\section{\revisedrone{Introduction}}\label{intro}
\revisionnoter{1}{Streamlined the introduction for improved clarity and flow}

\revisedrone{
Cryptography is essential for securing data in digital systems, with post-quantum cryptography (PQC) emerging as critical in response to quantum computing threats. PQC methods are resource-intensive due to larger key sizes—Dilithium2's signature is 9× larger than RSA-2048's \cite{Lyubashevsky2020-cc}—leading to significant performance and energy efficiency challenges \cite{nejatollahiPostQuantumLatticeBasedCryptography2019a,chenReportPostQuantumCryptography2016a,computersecuritydivisionSelectedAlgorithms20222017b}.

ASIC-based accelerators offer low latency but lack flexibility for diverse PQC algorithms \cite{verbauwhede24CircuitChallenges2015a,banerjeeEnergyEfficientConfigurableLattice2019e}. Our analysis reveals that PQC's primary bottleneck is limited cache bandwidth. Memory computing solutions address this through significant throughput and energy gains \cite{taylorProcessinginMemoryTechnologyMachine2022,shanbhagBenchmarkingInMemoryComputing2022a}. While off-chip designs expose data beyond chip boundaries, on-chip solutions keep computations secure within the processor, making them particularly suited for compute-intensive cryptographic operations.

\begin{figure}[tp]
\centerline{\includegraphics[width=5.0in]{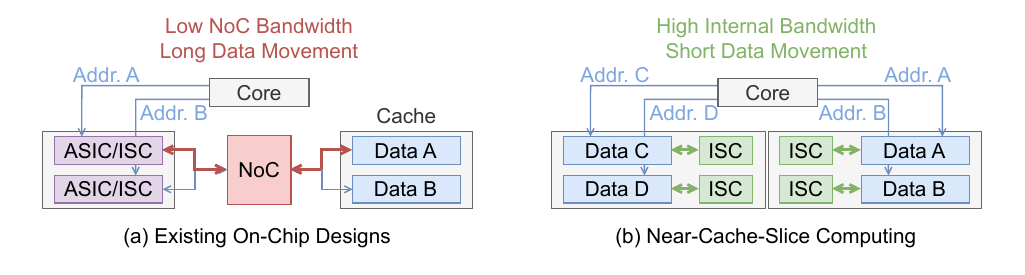}}
\caption{\revisedadd{Comparative illustration of data fetching paradigms. (a) Conventional on-chip designs with lower NoC bandwidth and extended data movement. (b) Our Near-Cache-Slice Computing achieving higher internal bandwidth via In-SRAM Computing.}\revisionnoteadd{Merged two figures into one for better visualization}}
\label{overview}
\end{figure}

State-of-the-art on-chip solutions employ either in-cache computing \cite{agaComputeCaches2017c,zhangRecryptorReconfigurableCryptographic2018a}, which repurposes existing caches but faces integration challenges, or near-cache computing \cite{reisIMCRYPTOInMemoryComputing2022a,chenEightCoreRISCVProcessor2022}, which adds dedicated processing elements but suffers from external bandwidth limitations (detailed comparison in Section \ref{ni-cache}).

Our analysis of cryptographic workloads reveals that input data typically fits within 64-byte cache blocks \cite{Lyubashevsky2020-cc,Avanzi2017-ge}, enabling our key insight: \textit{using the same virtual address for both data movement and computation}. This approach avoids the system integration challenges of in-cache computing while maintaining addressing transparency.

We propose \textit{Crypto-Near-Cache (CNC)}, a Near-Cache-Slice In-SRAM Computing design that combines in-cache computing's flexibility with near-cache computing's integration simplicity. As shown in Fig. \ref{overview}, CNC eliminates NoC bottlenecks by placing compute-enabled SRAM arrays adjacent to each cache slice, leveraging high internal bandwidth while supporting virtual addressing through ISA extensions. Unlike bit-serial approaches, CNC employs bit-parallel computing with flexible Computing Blocks (CBs) tailored to different algorithms' precision requirements, ensuring high performance for PQC kernels.

The contributions of this paper are:
1) Near-Cache-Slice Computing approach combining benefits of in-cache and near-cache designs (\S \ref{bg}).
2) CNC architecture with virtual addressing support and seamless cache integration (\S \ref{sec:arch}).
3) Algorithm-architecture co-design techniques for cryptographic acceleration (\S \ref{co-design}).
4) Comprehensive evaluation demonstrating significant energy and throughput improvements (\S \ref{sec:evaluation}).

}

\section{Motivation and Background} \label{bg}

This section provides a detailed categorization of in-cache and near-cache computing schemes, as illustrated in Fig. \ref{nearInCache} and summarized in \revised{Table \ref{table-cache-levels}}. We analyze their respective strengths and limitations to motivate our Near-Cache-Slice In-SRAM Computing approach.

\subsection{In-Cache \& Near-Cache Computing} \label{ni-cache}
\begin{figure*}[htp]
\centerline{\includegraphics[width=5.6in]{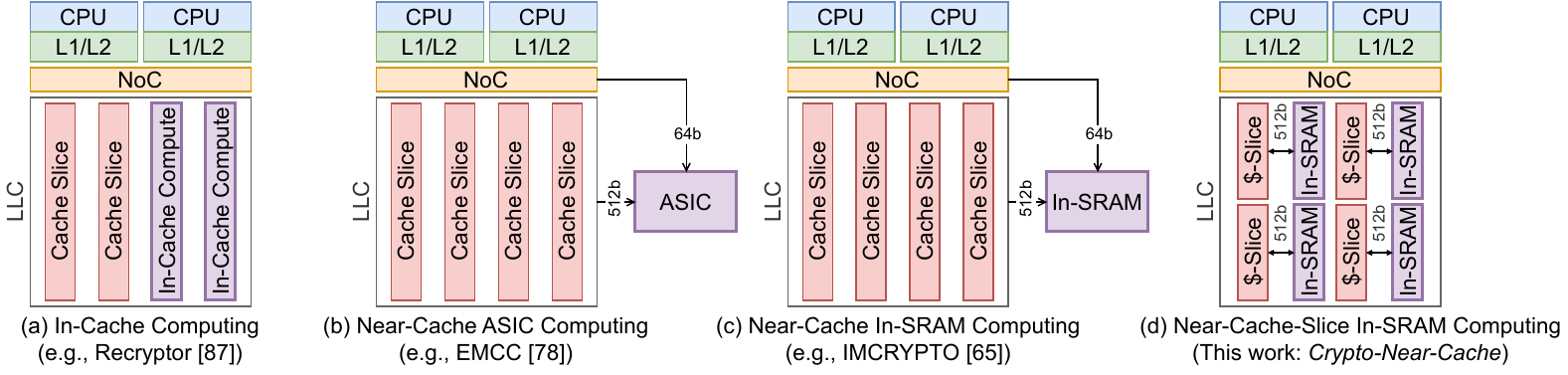}}
\caption{Design overview of (a) In-Cache Computing, (b) Near-Cache ASIC Computing, (c) Near-Cache In-SRAM Computing, (d) and Near-Cache-Slice In-SRAM Computing (This Work).
}
\label{nearInCache}
\end{figure*}

\textbf{In-Cache Computing:} 
In-Cache Computing, by virtue of its ability to directly compute on data within the cache, significantly reduces data movement. \hlc[white]{Its integration feasibility on device level has been commercially verified \mbox{\cite{noauthor_undated-tr}}.} Typically, as shown in Fig. \ref{nearInCache}(a), repurposing portions of the Last Level Cache into large vector processing units capable of bitline computing greatly enhances computational parallelism and energy efficiency. For instance, solutions like Recryptor \cite{zhangRecryptorReconfigurableCryptographic2018a} and Compute Cache \cite{agaComputeCaches2017c} employ a bit-parallel data layout to achieve cryptographic acceleration and general-purpose processing, respectively. Similarly, Neural Cache \cite{eckertNeuralCacheBitSerial2018} and Duality Cache \cite{fujikiDualityCacheData2019f} use a bit-serial data layout to support neural network acceleration and general-purpose processing.

The flexibility and generality of In-Cache Computing is rooted in the support for basic logic operations derived from In-SRAM Computing. This foundation allows it to accommodate a wide range of algorithms and even facilitates general-purpose computing.
A key characteristic of In-SRAM computing is that operands must share the same bitline, so that bitline computing can be enabled to perform XOR/AND/OR vector operations on operands \cite{jeloka28NmConfigurable2016b}.
This means each bit of different operands resides in a unique row but shares the same column within the SRAM array.
However, a significant challenge emerges when trying to integrate In-Cache Computing into modern computer systems, which often employ techniques like virtual addressing, set-associative cache, and address hashing.

\begin{figure}[tp]
\centering
\begin{minipage}[c]{2.4in}
    \centering
    \includegraphics[width=\linewidth]{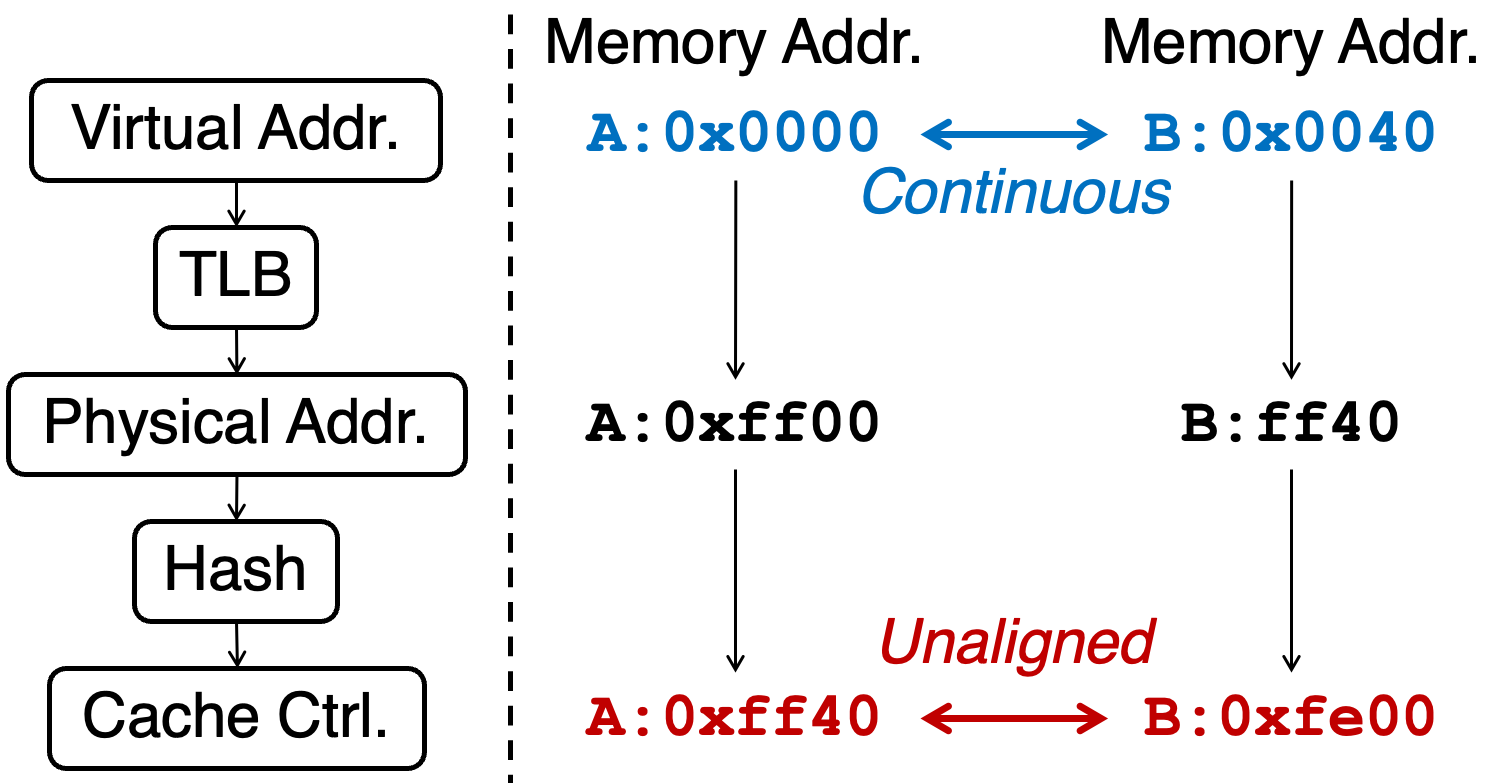}\\
    (a)
\end{minipage}
\hspace{0.3in}
\begin{minipage}[c]{2.2in}
    \centering
    \includegraphics[width=\linewidth]{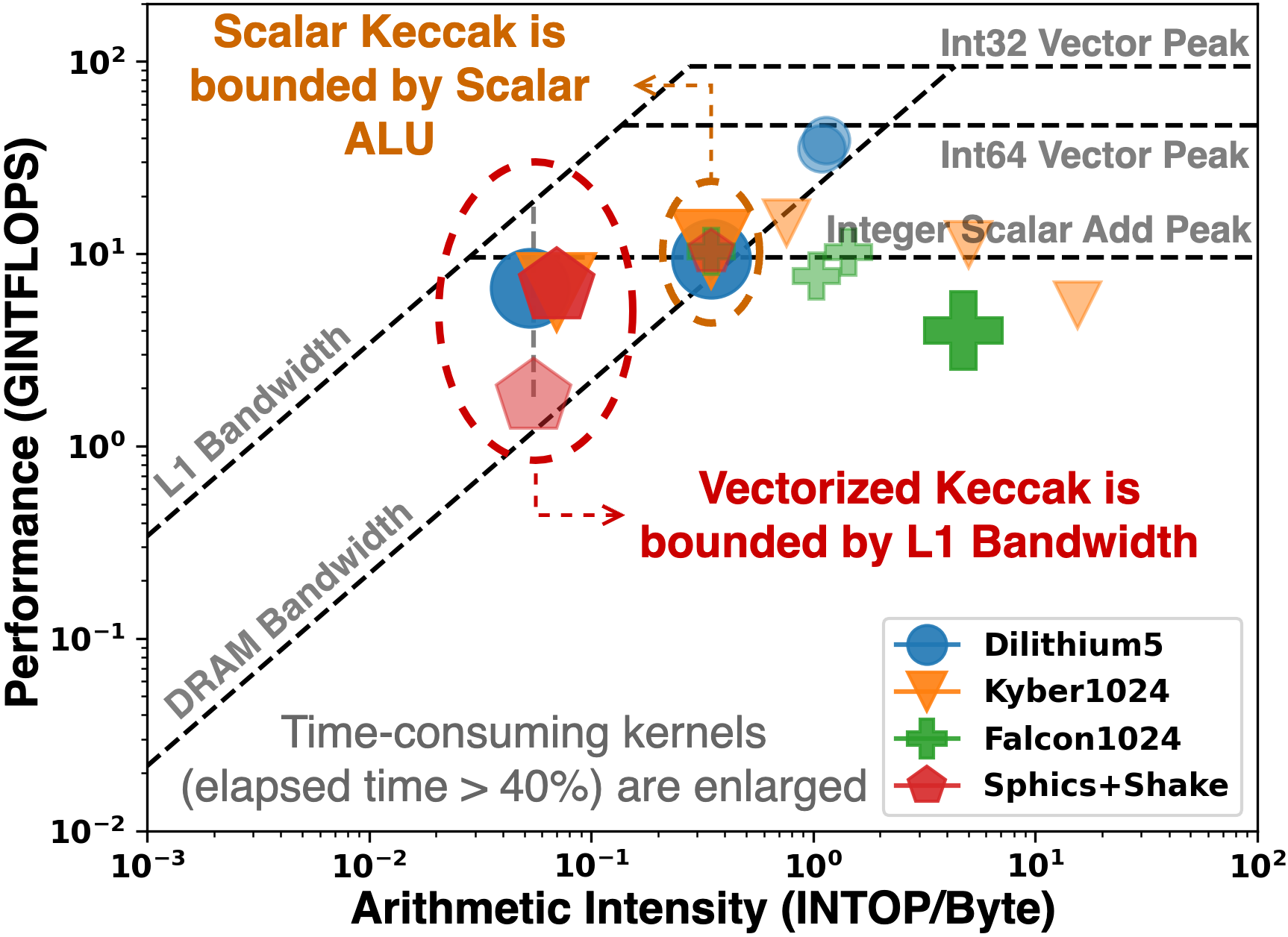}\\
    (b)
\end{minipage}
\caption{\revisedadd{(a) Process of address translation in modern computer systems and example of virtual address translation which makes the data layout uncontrollable in cache \cite{Yarom2015-ab}. (b) \hlc[white]{Roofline model for NIST-selected post-quantum cryptographic algorithms. Time-consuming kernels are enlarged.}}\revisionnoteadd{Merged two figures into one for better visualization}}
\label{feasability}
\end{figure}

Let's consider the deployment of virtual addressing which leads to an unpredictable data layout in the cache \cite{Yarom2015-ab}, as shown in Fig. \ref{feasability}. Assume that a program intends to write two consecutive 2-byte data units (\bera{A} and \bera{B}) into the memory one by one. As depicted in Fig. \ref{feasability}(a), the virtual address is converted to the physical address using the Translation Lookaside Buffer (TLB) 
This physical address is subsequently hashed to generate the final physical address that the cache controller uses to record the data. 
Despite \bera{A} and \bera{B} being located in successive memory locations from the programming view, post-TLB and hashing operations, their addresses get "randomly" mapped to different locations, as illustrated in Fig. \ref{feasability}(b).
While virtual addressing simplifies programming by offering a unified and continuous memory space, it also relinquishes control over the physical location of data in the cache and memory from the program's perspective and does not guarantee \hlc[white]{the} locality of operands and alignment of data. 
\hlc[white]{Although we can divide the cache into data cache and compute cache, retrieving data from the data cache for computation in the compute cache every time, the overhead of data movement can be as high as 90\% \mbox{\cite{Al-Hawaj2023-rr}}.}
As a result, implementing In-Cache Computing becomes a challenge in existing modern computer systems, unless there are significant modifications to the operating system \cite{agaComputeCaches2017c,zhangRecryptorReconfigurableCryptographic2018a} or considerable efforts on data transformation \cite{eckertNeuralCacheBitSerial2018,fujikiDualityCacheData2019f,wangInfinityStreamPortable2023}.

\textbf{Near-Cache ASIC/In-SRAM:}
To circumvent the system integration challenges associated with In-Cache Computing, Near-Cache Computing was introduced \cite{reisIMCRYPTOInMemoryComputing2022a,wangEagerMemoryCryptography2022,noriREDUCTKeepIt2021a,dharFReaCCacheFoldedlogic2020,chenEightCoreRISCVProcessor2022}. As its name suggests, Near-Cache Computing typically situates computing modules outside the Cache System, relying on the existing narrow 64-bit NoC or additional wide 512-bit data channels to facilitate data communication with the LLC, as illustrated in Fig. \ref{nearInCache}(b-c). 
As it merely requires the allocation of additional storage space, computing modules can be incorporated into the NoC or the cache system. Therefore, data movement can be facilitated by designating different address spaces.
The computing modules in Near-Cache Computing can take the form of ASICs (e.g., EMCC \cite{wangEagerMemoryCryptography2022}, REDUCT \cite{noriREDUCTKeepIt2021a}) to efficiently support a subset of computations, or they can be constructed as arrays utilizing In-SRAM Computing (e.g., IMCRYPTO \cite{reisIMCRYPTOInMemoryComputing2022a}) to flexibly, generally and efficiently support a wide range of computations, even to the extent of general-purpose computing with large vector processing units (repurposed SRAM wordlines).
Although this design circumvents the pain of system integration, its effectiveness is still constrained by the external bandwidth of the cache (at most 512-bit, necessitating modifications to the data/control paths), which is hard to enable multiple computing modules.

\textbf{Near-Cache-Slice In-SRAM:} 
In light of these challenges, our proposed Near-Cache-Slice In-SRAM Computing, also referred to as Crypto-Near-Cache, combines the flexibility and large vector processing of In-Cache Computing with the ease of integration of Near-Cache Computing, as shown in Fig. \ref{nearInCache}(d). 
Specifically, we place an array using In-SRAM Computing near each cache slice to leverage both the large internal bandwidth and the flexibility/generality of In-SRAM Computing with large vector processing units (repurposed SRAM wordlines).
Compared to In-Cache Computing, Near-Cache-Slice Computing is easier to be integrated into systems. This is because the compute-enabled SRAM array in Near-Cache-Slice computing uses physical addresses within its control module. Therefore, the control module can finely manage where data is written in compliance with the requirements of bitline computing and data alignment. This avoids the situation that can occur with In-Cache Computing systems where the physical address of data cannot be controlled due to the use of virtual addresses.
This design offers a general, flexible, high-throughput, and easily integrable solution for accelerating cryptographic computations, and even extends to general-purpose computations.

\subsection{Overview of Cryptography Workloads}
Cryptographic computations play a critical role in ensuring secure communication in modern society. For instance, key exchange algorithms like RSA facilitate the creation of shared keys between parties to prevent leakage \cite{rivestMethodObtainingDigital1978a}. Block encryption algorithms, such as AES, help to maintain confidentiality for various types of data, including data streams, by utilizing shared keys \cite{daemenRijndaelBlockCipher}. As these algorithms have been in use for many years and are widely used, many hardware platforms feature dedicated engines to accelerate them \cite{IntelAdvancedEncryption}.

Moreover, the rise of quantum computing has emerged as a significant threat to public-key cryptography. The National Institute of Standards and Technology (NIST) has selected four post-quantum algorithms in 2022 \cite{computersecuritydivisionSelectedAlgorithms20222017b}, Crystals-Kyber \cite{bosCRYSTALSKyberCCASecure2018a}, Crystals-Dilithium \cite{CRYSTALSDilithiumLatticeBasedDigital}, Falcon \cite{Falcona}, and Sphincs+ \cite{SPHINCSSignatureFramework}.
However, executing these algorithms effectively on existing computing systems
poses computational challenges \cite{kumarSecuringFutureInternet2022b}, stemming from the utilization of long public key and signature, and complex primary building block kernels such as Keccak-1600 for security assurance and number theoretic transform (NTT) for lattice-based cryptography.

\section{Rationality behind \textit{Crypto-Near-Cache} from Application Perspective} \label{where-how}

In this section, we explain why \textit{Crypto-Near-Cache} fits compute-intensive workloads, particularly post-quantum cryptographic (PQC) tasks. We discuss key application requirements for on-chip architecture---security, energy, performance, and flexibility---and offer corresponding takeaways.

\textbf{Security.}
Protecting user data and privacy is paramount in cryptographic computation. Due to the risk of compromised hypervisors in cloud environments and the vulnerability of IoT devices, a widely accepted principle is that ``only the chip itself is considered the TCB (Trusted Computing Base), and no sensitive plaintext data is stored off-chip'' \cite{IntelTrustDomain, AMDSecureEncrypted, ltdTrustZoneCortexMArm}. Although attestation \cite{Volos2018-ca, Chen2019-og} can extend the TCB to peripherals, it increases overhead and the attack surface \cite{Mai2023-sx, Ivanov2023-qa}. Alternative privacy-preserving technologies (e.g., homomorphic encryption \cite{Samardzic2021-fi} or multi-party computation \cite{Xiong2022-dk}) impose even higher performance costs than TEEs. 
\textit{\textbf{Takeaway I:} Cryptographic workloads must run on-chip for security. Off-chip accelerators (e.g., in-NVM \cite{xieSecuringEmergingNonvolatile2018a, liuFeCryptoInstructionSet2022} or in-ReRAM \cite{nejatollahiCryptoPIMInmemoryAcceleration2020a, parkRMNTTRRAMBasedComputeinMemory2022b}) are considered insecure.}

\textbf{Energy Consumption.}
Rising sustainability demands and battery-powered IoT devices make energy usage critical. Using Sniper \cite{carlson2014aeohmcm} to profile PQC algorithms shows that core and cache operations dominate energy costs. By performing computation directly in the SRAM array with SIMD-like parallelism, in-SRAM computing reduces instruction overhead and eliminates most data-movement energy. Its analog-like operations consume less power than digital logic, improving efficiency for power-sensitive devices.
\textit{\textbf{Takeaway II:} In-SRAM computing with SIMD-like execution minimizes instruction overhead and data transfers, significantly saving energy for cryptographic tasks.}

\textbf{Performance.}
A roofline model via Intel Advisor \cite{marquesPerformanceAnalysisCacheAware2017} reveals that PQC workloads are compute-intensive, as shown in Fig. \ref{feasability}(b). Time-consuming vector kernels saturate L1 bandwidth, whereas scalar kernels max out scalar ALUs. Off-chip bandwidth is less of a limiting factor here; higher on-chip bandwidth improves vector-kernel performance, and increased vector hardware accelerates scalar kernels.
\textit{\textbf{Takeaway III:} Cryptographic workloads demand higher data bandwidth and additional vector units to overcome on-chip compute bottlenecks.}

\textbf{Flexibility/Generality.}
Each cryptographic algorithm has multiple configurations (e.g., Dilithium-2/3/5 \cite{Lyubashevsky2020-cc}, Kyber-512/768/1024 \cite{Avanzi2017-ge}), and cryptography evolves quickly \cite{zhaoSurveyDifferentialPrivacy2022, braunMOTIONFrameworkMixedProtocol2022, albadawiOpenFHEOpenSourceFully2022}. Future primitives like lightweight cryptography \cite{ghoshLightweightPostQuantumSecureDigital2019, kumarSecuringFutureInternet2022b, mustafaLightweightPostQuantumLatticeBased2020} will further diversify requirements. Repeatedly designing new ASICs is impractical.
\textit{\textbf{Takeaway IV:} Flexible, general-purpose solutions are essential for ever-changing cryptographic algorithms, making fixed ASIC solutions less suitable \cite{banerjeeSapphireConfigurableCryptoProcessor2019b, chenReportPostQuantumCryptography2016a, songLEIA05mm2140mW2018a}.}

\begin{table}[tp]
\centering
\caption{\revisedrtwo{Comparison of Near-Cache Computing Levels.}}
\scalebox{0.85}{
\begin{tabular}{>{\centering\arraybackslash}p{0.5in}cccccc}
\toprule
\revisedrtwo{Level} & \revisedrtwo{Bits per} & \revisedrtwo{Coherence \&} & \revisedrtwo{Bits} & \revisedrtwo{Integration} & \revisedrtwo{Address} & \revisedrtwo{Security} \\
      & \revisedrtwo{Cycle} & \revisedrtwo{Consistency} & \revisedrtwo{Consolidation} & \revisedrtwo{Complexity} & \revisedrtwo{Translation} & \revisedrtwo{Isolation} \\
\midrule
Cache & \texttt{n} & \cmark & \xmark & Low & \cmark & \cmark \\
\revisedrtwo{Slice} & \revisedrtwo{\texttt{nm}} & \revisedrtwo{\cmark} & \revisedrtwo{\xmark} & \revisedrtwo{Low} & \revisedrtwo{\cmark} & \revisedrtwo{\cmark} \\
Way & \texttt{nmk} & \xmark & \xmark & High & \xmark & Partial \\
Bank & \texttt{nmk} & \xmark & \cmark & High & \xmark & \cmark \\
\bottomrule
\end{tabular}
}
\label{table-cache-levels}

\begin{center}
\parbox{0.9\linewidth}{
\small \textbf{Note:} Assume cache external bandwidth of $n$ bits/cycle, with $m$ slices, $k$ ways per slice. \textbf{Coherence/consistency:} compatibility with existing cache protocols. \textbf{Bits consolidation:} need for cross-array data movement. \textbf{Integration complexity:} difficulty of incorporating into existing hierarchies. \textbf{Address translation:} support for virtual addressing. \textbf{Security isolation:} computation isolation from main data path. \revisionnoter{2}{Added detailed comparison table to clarify different near-cache design approaches and justify our slice-level choice}
}
\end{center}
\vspace{-2em}
\end{table}

\textbf{\revisedrtwo{Different Levels among Near-Cache Solutions.}}
\revisionnoter{2}{Added comprehensive analysis of different near-cache solution levels}
\revisedrtwo{Table~\ref{table-cache-levels} compares various near-cache designs across six key dimensions. The Near-Cache-Slice approach achieves optimal balance: it provides high internal bandwidth (nm\,bits/cycle), maintains coherence and consistency compatibility, supports address translation for virtual memory, and ensures security isolation—all with low integration complexity. Unlike way-level and bank-level designs that sacrifice coherence or require complex bit consolidation, the slice-level approach avoids these limitations while delivering comparable bandwidth. This combination is particularly beneficial for PQC workloads, which demand both high-throughput computation and secure on-chip data processing.}

\section{Crypto-Near-Cache Architecture} \label{sec:arch}

\subsection{Crypto-Near-Cache Overview}

\begin{figure}[htp]
\centering
\begin{minipage}[c]{0.5\textwidth}
\centering
\includegraphics[width=\textwidth]{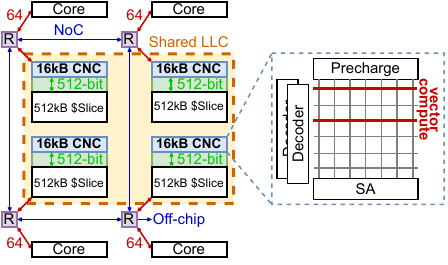}\\
(a)
\end{minipage}
\hfill
\begin{minipage}[c]{0.48\textwidth}
\centering
\includegraphics[width=\textwidth]{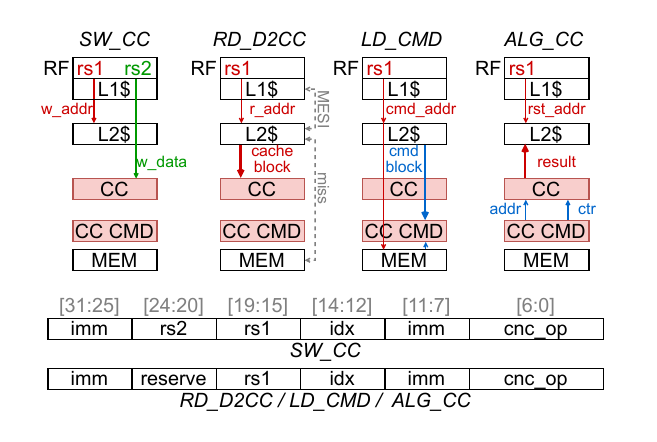}\\
(b)
\end{minipage}
\caption{\revisedadd{(a) Overview of Crypto-Near-Cache-enabled system. (b) Crypto-Near-Cache ISA extensions and diagrams of how CNC-enabled system supports CNC instructions.}\revisionnoteadd{Merged two figures into one for better visualization}}
\label{fig:cnc-system}
\end{figure}

The Crypto-Near-Cache-enabled system, illustrated in Fig. \ref{fig:cnc-system}(a), integrates a CNC module into each slice of the shared last-level cache (L2 in embedded devices). This module comprises an SRAM array with bitline computing capabilities, optimised to accelerate cryptographic algorithms. CNC units can concurrently execute distinct cryptographic algorithms across different cache slices, leveraging high internal cache slice bandwidth for efficient data fetching without disrupting existing read/write mechanisms. Through simultaneous activation of two wordlines, each subarray row serves as a large vector for computation, enabling the SRAM array to dual-function as both a computational and storage unit.

\subsection{Instruction Set Architecture (ISA) Extension} \label{sec:isa}

\revisedrtwo{This subsection presents the formal specification of CNC's ISA extension. We extend RISC-V's S-type instruction format to support four CNC-specific instructions that enable efficient cryptographic computation.}

\subsubsection{\revisedrtwo{Formal Instruction Specification}}

\revisedrtwo{Table \ref{tab:isa-spec} provides the complete specification of CNC instructions, including their encoding, semantics, and timing characteristics.} \revisionnoter{2}{Added formal ISA specification table addressing R-2}

\begin{table}[htp]
\centering
\caption{\revisedrtwo{CNC Instruction Set Extension Specification}}
\label{tab:isa-spec}
\scalebox{0.55}{
\begin{tabular}{|l|l|l|l|c|}
\hline
\textbf{Instruction} & \textbf{Encoding (32-bit)} & \textbf{Syntax} & \textbf{Operation} & \textbf{Cycles} \\
\hline
SW\_CNC & \texttt{imm[11:5]|rs2|rs1|010|imm[4:0]|0101011} & \texttt{sw\_cnc rs2, imm(rs1)} & \makecell[l]{CNC[addr] $\leftarrow$ GPR[rs2]\\addr = GPR[rs1] + sext(imm)\\Bypass L1 cache} & 1 \\
\hline
RD\_D2CNC & \texttt{imm[11:5]|5'b0|rs1|011|imm[4:0]|0101011} & \texttt{rd\_d2cnc imm(rs1)} & \makecell[l]{CNC[0:511] $\leftarrow$ Cache[block\_addr]\\block\_addr = align64(GPR[rs1] + sext(imm))\\Transfer entire cache block} & 2 \\
\hline
LD\_CMD & \texttt{imm[11:5]|5'b0|rs1|100|imm[4:0]|0101011} & \texttt{ld\_cmd imm(rs1)} & \makecell[l]{CMD\_ARRAY $\leftarrow$ MEM[addr:addr+size]\\addr = GPR[rs1] + sext(imm)\\size determined by algorithm} & varies \\
\hline
ALG\_CNC & \texttt{imm[11:5]|alg[4:0]|rs1|101|imm[4:0]|0101011} & \texttt{alg\_cnc imm(rs1)} & \makecell[l]{Execute algorithm specified by alg[4:0]\\MEM[result\_addr] $\leftarrow$ CNC\_result\\result\_addr = GPR[rs1] + sext(imm)} & \makecell[c]{alg-\\specific} \\
\hline
\end{tabular}
}
\end{table}

\subsubsection{\revisedrtwo{Instruction Semantics and Constraints}}\label{sec:isa-semantics}
\revisionnoter{2}{Added precise instruction semantics}

\revisedrtwo{\textbf{SW\_CNC (Store Word to CNC):} This instruction bypasses the L1 cache and directly writes a 32-bit word from a general-purpose register to the CNC array. The virtual address is computed as GPR[rs1] + sign-extended immediate. This is typically used to load small parameters such as AES counters (32-bit) or random seeds for Kyber/Dilithium. No alignment requirement is imposed on the address, cache state remains unaffected, and standard TLB translation applies for virtual address handling. 

\textbf{RD\_D2CNC (Read Data to CNC):} Transfers an entire 64-byte cache block from the L2 data cache to the CNC array using the high-bandwidth internal cache interface. The block address must be 64-byte aligned (enforced via \texttt{aligned\_alloc()}). This instruction achieves 512 bits transfer in 2 cycles while maintaining cache coherence and triggering miss handling when needed.

\textbf{LD\_CMD (Load Commands):} Loads pre-computed control commands from memory into the CNC command array. Commands define the sequence of operations for cryptographic algorithms. Command sizes are algorithm-specific (e.g., 5.9k commands for AES-128, 7.0k for Keccak-1600), can be cached for reuse across multiple operations, and access memory through standard cache hierarchy traversal.

\textbf{ALG\_CNC (Algorithm Execution):} Executes a specific cryptographic algorithm using the loaded commands. The \texttt{alg[4:0]} field specifies the algorithm and its variant. Different parameter sets (e.g., AES-128/256, Keccak-1600, NTT-256, Kyber512/768/1024, Dilithium2/3/5) are assigned unique 5-bit encodings, supporting up to 32 algorithm variants. The execution time varies significantly based on algorithm complexity, ranging from tens of thousands of cycles for basic primitives to hundreds of millions for complete PQC protocols (detailed performance analysis in Section~\ref{sec:evaluation}).}

\subsubsection{\revisedrone{Virtual Address Handling Mechanism}}\label{sec:virtual-address}
\revisionnoter{1}{Streamlined description} 

\revisedrone{CNC instructions use virtual addresses, enabling seamless integration with existing memory management. The translation process follows standard pipeline flow: (1) compute virtual address using GPR[rs1] + sign-extended immediate; (2) translate through data TLB in Memory stage; (3) use destination hash module to select cache slice; (4) access the appropriate CNC array. This approach ensures \textbf{transparency} (applications unaware of physical placement), \textbf{security} (respects page permissions), and \textbf{compatibility} (works with OS features like demand paging and copy-on-write) while maintaining cache coherence.} 

\subsubsection{\revisedrtwo{Compiler Integration}}  \label{sec:compiler-integration}
\revisionnoter{2}{Added compiler integration details}

\revisedrtwo{Custom instructions in CNC are integrated into compilers through LLVM intrinsics. Users interact with a Post-Quantum Cryptography (PQC) library, marking functions for CNC acceleration via pragma directives:} 

\vspace{2.1em}

\begin{lstlisting}[language=C, caption={\revisedrtwo{CNC acceleration via pragma directive}}, label={lst:cnc-pragma}]
#pragma cnc_accelerate
void kyber512_encrypt(uint8_t *ct, const uint8_t *pk, 
                      const uint8_t *msg, const uint8_t *coins) {
    // Function implementation
}
\end{lstlisting}

This instructs the compiler to generate LLVM intrinsic calls:
\begin{lstlisting}[language=LLVM, caption={\revisedrtwo{LLVM intrinsic for CNC acceleration}}, label={lst:llvm-intrinsic}]
llvm.call @pim_kyber512(%input_vector, %output_vector, %size) 
    : (!llvm.ptr, !llvm.ptr, i32) -> ()
\end{lstlisting}

\revisedrtwo{At runtime, these intrinsics are translated into CNC ISA instructions (SW\_CNC, RD\_D2CNC, LD\_CMD, ALG\_CNC) as described above. The complete compiler toolchain will be open-sourced to facilitate community validation and adoption.

Based on our proposed ISA extension, the programming model is flexible. CNC can be utilized either through a library-based approach (similar to CUDA \cite{Sanders2010-pu}) or by using inline compilation methods like \textit{pragma}.
}

\subsubsection{\revisedrtwo{Cryptographic Algorithm Mapping}}
\revisionnoter{2}{Added code examples for AES-128 and NTT operations}

\begin{figure}[htp]
\centering
\begin{minipage}[t]{0.48\textwidth}
\textbf{Example 1: AES-128 Encryption}
\begin{lstlisting}[language=CNC-Assembly, caption={\revisedrtwo{AES-128 encryption using CNC instructions}}, label={lst:aes-example}]
# Load AES key schedule commands
li      t0, aes_keysched_cmds
ld_cmd  0(t0)
li      t1, key_addr  # Load key into CNC
rd_d2cnc 0(t1)
# Execute key expansion
li      t2, round_keys_addr 
aes_cnc 0(t2)
li      t0, aes_round_cmds  # Load commands
ld_cmd  0(t0)
li      t1, plaintext_addr  # Load plaintext
rd_d2cnc 0(t1)
# Execute AES encryption
li      t3, ciphertext_addr
aes_cnc 0(t3)
\end{lstlisting}
\end{minipage}
\hfill
\begin{minipage}[t]{0.48\textwidth}
\textbf{Example 2: NTT for Kyber512}
\begin{lstlisting}[language=CNC-Assembly, caption={\revisedrtwo{NTT computation for Kyber512}}, label={lst:ntt-example}]
# Load NTT butterfly commands
li      t0, ntt_cmds
ld_cmd  0(t0)

# Load polynomial coefficients
# (4 cache blocks)
li      t1, poly_addr
rd_d2cnc 0(t1)    # Coeff 0-15
rd_d2cnc 64(t1)   # Coeff 16-31
rd_d2cnc 128(t1)  # Coeff 32-47
rd_d2cnc 192(t1)  # Coeff 48-63

# Execute NTT transform
li      t2, ntt_result_addr
ntt_cnc 0(t2)
\end{lstlisting}
\end{minipage}
\end{figure}

\revisedrtwo{To illustrate how CNC instructions enable cryptographic computation, Listing~\ref{lst:aes-example} and Listing~\ref{lst:ntt-example} provide concrete examples for AES-128 encryption and NTT computation respectively.

The \texttt{aes\_cnc} instruction behavior depends on previously loaded commands—key expansion for the first invocation, encryption for the second. Similarly, NTT uses \texttt{ld\_cmd} to define butterfly patterns and leverages high cache bandwidth (512 bits in 2 cycles) for efficient coefficient transfer. These examples demonstrate how CNC decomposes complex cryptographic operations into simple instruction sequences while exploiting parallelism.
}

\subsection{Core and LLC Datapath Design} \label{sec:datapath}

\begin{figure*}[htp]
\centerline{\includegraphics[width=5.8in]{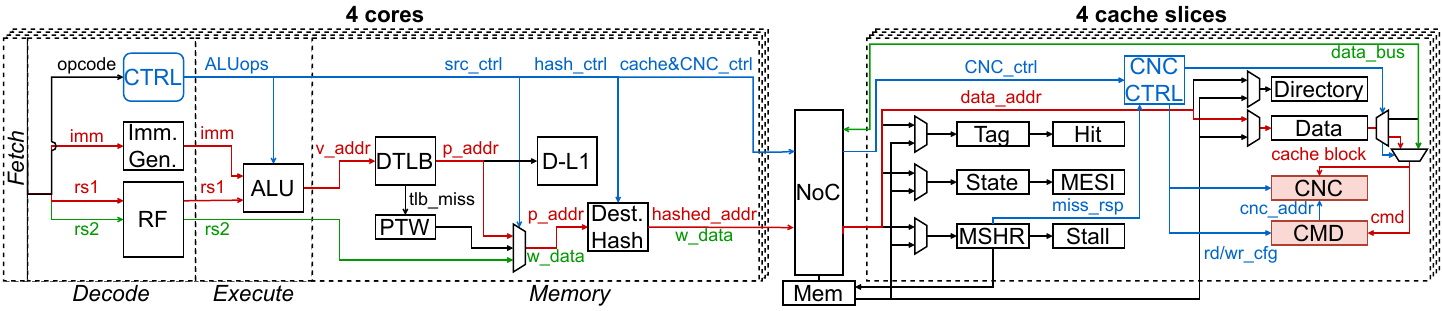}}
\caption{CNC-based core and cache datapath including CNC array, command array, and control module, are highlighted.
\vspace{-1em}
}
\label{datapath}

\end{figure*}

This section explores the core and cache datapath design of the CNC, utilizing a RISC-V pipeline design for the core \cite{pattersonComputerOrganizationDesigna} and an OpenPiton-based \cite{chenEightCoreRISCVProcessor2022,balkindOpenPitonOpenSource2016,mckeownPitonManycoreProcessor2017} cache structure. The flexible system integrates four cores and cache slices and can be scaled to fit requirements. While the main focus is integrating CNC into existing modern computer systems, for microcontroller-only IoT devices, a tailored cache design can be achieved by removing cache coherence-related modules.

\textbf{Core datapath:} Supporting CNC-related instructions necessitates pipeline modifications, which involve extending the capabilities of core's control module and managing the destination hash module for accurate cache slice access. For single-register instructions (e.g., \textit{RD\_D2CNC}, \textit{LD\_CMD}, and \textit{ALG\_CNC}), the Program Counter (PC) retrieves the instruction from the I-Cache before passing it to the \textit{Decode} stage. Here, the CTRL module decodes the opcode, generating control signals for subsequent pipeline stages, cache, and CNC operations (Fig. \ref{datapath}). The \textit{imm} and \textit{rs1} parameters proceed to the \textit{Execute} stage, where the virtual address for the target data is generated.
Note that the address must be aligned to a cache block for \textit{RD\_D2CNC}, guaranteed with \texttt{aligned\_alloc()}.

In the \textit{Memory} stage, the data TLB translates the virtual address into a physical address (PA). In case of a TLB miss, the memory system conducts a page table walk. Upon address translation completion, the PA bypasses the L1 cache, avoiding unnecessary access, and proceeds directly to higher memory levels. The control module generates signals to manage the multiplexer for PA selection, while the destination hash module performs a hash operation, producing the final PA for the cache. Lastly, the PA and associated control signal (cache\&CNC\_ctrl) are bundled and forwarded to the NoC.

For \textit{SW\_CNC} instructions with two register fields, the destination hash module's control differs slightly from previous instructions. During \textit{SW\_CNC} execution, the module first calculates routing information based on the instruction's \textit{rs1} field. Subsequently, when packaging the write data (\textit{w\_data}) fetched by \textit{rs1}, the hash module incorporates the same routing information, ensuring correct cache slice transmission.

By ensuring that all instructions (\textit{SW\_CNC}, \textit{RD\_D2CNC}, \textit{LD\_CMD}, \textit{ALG\_CNC}) use the same read/write address, we can guarantee that they are operating on the same data. Given that our target application is cryptographic computation, the input generally does not exceed one cache block (64 bytes). If the input exceeds one cache block, we first read the initial cache block into the CNC array, then overwrite any data exceeding one cache block in the same cache position before reading it into the CNC array.


\textbf{Cache datapath:} 
Upon receiving requests from the core, the NoC transmits request packages to one of the cache slices via routers. Each cache slice features a CNC array, a command array for control operations, and an auxiliary control module. For the \textit{SW\_CNC} write instruction, \textit{w\_data} on the \textit{data\_bus} is written to the CNC through MUX selection. The control module incrementally provides written addresses for the CNC, starting from a predefined number (e.g., 0).

For the \textit{RD\_D2CNC} instruction, \textit{data\_addr} is initially checked in the Tag array. If a cache miss is identified, the memory request is directed to the main memory and managed by the miss status holding register (MSHR). Once the miss is resolved, the MSHR notifies the control module. The directory array then examines \textit{data\_addr}, and in case of a coherence miss, forwards the request to other cache components. If neither cache nor coherence misses occur, \textit{data\_addr} is sent to the data array, while the control module generates a corresponding read signal. The read signal retrieves the entire cache block, and the control module enables writing to the CNC. Consequently, data read from the data cache is directly written to the CNC, allowing cache block read and write operations within two adjacent cycles.

The \textit{LD\_CMD} instruction follows the same tag and coherence check process as \textit{RD\_D2CNC}. After resolving cache and coherence misses, commands are written from the data cache to the command array via MUX. For instructions executing specific algorithms (e.g., \textit{AES\_CNC}, \textit{Keccak\_CNC}, \textit{NTT\_CNC}, \textit{Kyber\_CNC}, or \textit{Dilithium\_CNC}), the control module sequentially reads commands from the command array, connecting each to the CNC's control logic, which allows the CNC to perform different algorithms based on logic operations. Upon algorithm completion, the result is written into the data array according to the \textit{data\_addr} received from the NoC.

\subsection{Crypto-Near-Cache Array Design}\label{cnc-design}

\begin{figure}[tp]
\centering
\begin{minipage}[c]{0.48\textwidth}
\centering
\includegraphics[width=\textwidth]{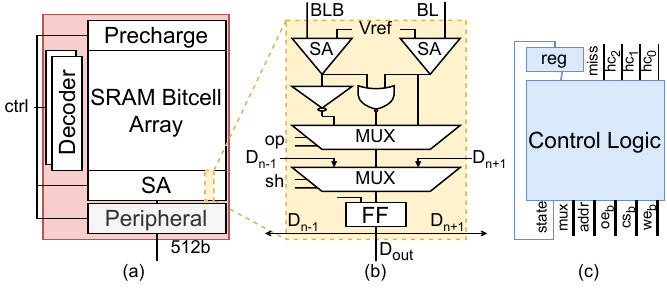}
\end{minipage}
\hfill
\begin{minipage}[c]{0.51\textwidth}
\centering
\includegraphics[width=\textwidth]{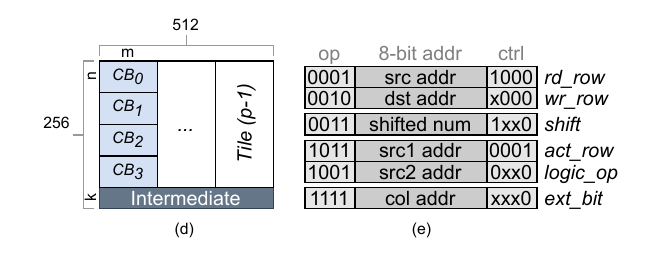}
\end{minipage}
\caption{(a) Diagram of Crypto-Near-Cache array. (b) Design of sense amplifier to support OR/XOR/AND and 1-bit left/right shift. (c) Diagram of Crypto-Near-Cache control module. (d) Computing Blocks (CBs) and tiles to support various cryptographic algorithms in Crypto-Near-Cache. (e) Formats of CNC control commands.}
\label{fig:cnc-arch-mapping}
\end{figure}

This section introduces the CNC array design. As shown in Fig. \ref{fig:cnc-arch-mapping}(a), CNC consists of an SRAM array with computing capabilities. The SRAM array employs two decoders for the simultaneous activation of two wordlines, enabling bitline computing. 
 
Our design uses a 256-row by 512-column SRAM array. 
To support various algorithms with different data widths, we employ Computing Blocks (CBs) as fundamental computational units, illustrated in Fig. \ref{fig:cnc-arch-mapping}(d). Each CB consists of \textit{n} rows and \textit{m} columns of SRAM cells, with dimensions varying based on algorithm requirements (e.g., \textit{n=4} and \textit{m=32} for AES-128/256, \textit{n=64} and \textit{m=25} for Keccak-1600, \textit{n=128} and \textit{m=16} for NTT-128, and \textit{n=256} and \textit{m=16} for NTT-256 and larger polynomial operations). 
The CNC uses \textit{k} rows for intermediate \hlc[white]{variables}, with \textit{k} flexibly adjusted based on application needs. Our analysis indicates that setting \textit{k} to 6 satisfies most algorithm requirements.
Furthermore, CBs that share the same set of bitlines form a tile, resulting in $p=\left \lfloor 512/m\right \rfloor$ tiles for a 512-column array. CBs that share the same wordline in different tiles can execute in parallel. For operations requiring data widths exceeding a single CB (e.g., NTT-512 or larger), multiple tiles within the same row collaborate by sharing wordlines, enabling seamless processing of extended data structures.

Fig. \ref{fig:cnc-arch-mapping}(b) depicts the sense amplifier (SA) design, which supports OR/XOR/AND and 1-bit left/right shift operations by incorporating a NOT and NOR gate, two 4:1 MUXes, and a flip-flop (less than 2\% overhead). {BL is bitline and BLB is bitline bar.} The MUX control signals are generated by the control module in Fig. \ref{fig:cnc-arch-mapping}(c), comprising a finite state machine that governs the data array, CNC, and command array read/write operations. Inputs include CNC signals from \textit{CNC\_ctrl} and \textit{miss} response from MSHR. The 3-bit CNC signal accommodates up to five algorithms and \textit{SW\_CNC}, \textit{RD\_D2CNC}, and \textit{LD\_CMD} data transfer instructions. The \textit{miss} signal indicates cache misses, causing the control logic to pause until data is fetched from memory. The \textit{web}, \textit{csb}, and \textit{oeb} signals, generated by the control logic, manage read/write operations for different arrays, while \textit{addr} signifies the current read/write address.

\subsection{\revisedrtwo{Detailed Micro-architecture}} \label{sec:microarch}

\revisedrtwo{This subsection provides comprehensive micro-architectural details of the CNC array, including internal organization and Computing Block (CB) mapping to cryptographic algorithms.} \revisionnoter{2}{Added detailed micro-architecture section addressing R-2}

\subsubsection{\revisedrtwo{CNC Array Internal Organization}} \label{sec:cnc-array-internal-organization}
\revisionnoter{2}{Added detailed micro-architecture description addressing R-2}

\revisedrtwo{The CNC array is physically implemented as a unified 256x512 SRAM array with bitline computing capabilities. Fig. \ref{fig:cnc-arch-mapping}(a) illustrates this structure, which incorporates dual decoders enabling simultaneous activation of two wordlines for bitline computing operations. 

Computing Blocks (CBs) are logical abstractions overlaid on this physical array to facilitate algorithm-specific operations. Each CB logically groups \textit{n} rows and \textit{m} columns, with dimensions dynamically configured based on the executing algorithm. This logical partitioning allows the same physical array to be reconfigured for different cryptographic workloads without hardware modifications. CBs sharing the same bitlines form logical tiles, enabling parallel execution across tiles.

The internal data flow follows a three-stage pipeline: (1) \textbf{Data Loading Stage:} Input data is loaded into designated CBs through the 512-bit data bus, with addresses generated by the control module. The loading process utilizes row-major ordering for sequential algorithms and customized patterns for parallel operations. (2) \textbf{Computation Stage:} Multiple CBs execute in parallel, performing logic operations (AND/OR/XOR) and shifts based on pre-loaded commands. The sense amplifiers (SAs) in Fig. \ref{fig:cnc-arch-mapping}(b) enable these operations with minimal overhead (<2\% area increase). (3) \textbf{Result Writeback Stage:} Computed results are written back to designated rows for subsequent operations or final output through the data bus.

The array includes 6 reserved rows for intermediate variables, enabling complex multi-step computations without external memory access. This design choice balances area efficiency with computational flexibility, as our analysis shows 6 rows suffice for all evaluated cryptographic algorithms.}

\subsubsection{\revisedrtwo{Computing Block Mapping to Algorithms}} \label{sec:computing-block-mapping-to-algorithms}

\revisedrtwo{The unified physical CNC array is logically reconfigured into different CB arrangements for each cryptographic algorithm to maximize parallelism and minimize data movement. Table \ref{tab:cb-mapping} details how the same 256x512 physical array is logically partitioned for different algorithms:} \revisionnoter{2}{Added CB mapping table demonstrating flexibility}

\added{
\begin{table}[h]
\centering
\caption{\revisedrtwo{Computing Block Configuration for Cryptographic Algorithms}}
\label{tab:cb-mapping}
\scalebox{0.75}{
\begin{tabular}{|l|c|c|c|c|c|}
\hline
\textbf{Algorithm} & \textbf{CB Size (nxm)} & \textbf{Tiles} & \textbf{Parallel CBs} & \textbf{Data Width} & \textbf{Utilization} \\
\hline
AES-128 & 4x32 & 16 & 16 & 128-bit & 100\% \\
AES-256 & 4x32 & 16 & 16 & 256-bit & 100\% \\
Keccak-1600 & 64x25 & 20 & 1 & 1600-bit & 100\% \\
NTT-128 & 128x16 & 32 & 2 & 2048-bit & 100\% \\
NTT-256 & 256x16 & 16 & 1 & 4096-bit & 100\% \\
Kyber512 & 256x16 & 16 & 1 & 4096-bit & 100\% \\
Dilithium2 & 256x16 & 16 & 1 & 4096-bit & 100\% \\
\hline
\end{tabular}
}
\end{table}
}

\revisedrtwo{Note: For NTT operations exceeding 256 points, multiple tiles located in the same row collaborate to process the larger polynomial. For instance, NTT-512 utilizes 2 tiles (2×256×16) sharing the same wordlines, while NTT-1024 uses 4 tiles, enabling efficient parallel computation across the extended coefficient array. \revisionnoter{2}{Demonstrated extensibility for larger operations}

\textbf{AES Mapping:} For both AES-128 and AES-256, the physical array is logically organized as 4x32 CBs, enabling 16 parallel AES operations across 16 logical tiles. The consistent CB size for both AES variants simplifies hardware implementation. The S-box operations utilize Galois Field arithmetic distributed across CB rows, eliminating lookup tables. Round keys are pre-computed and stored in reserved rows, accessed via single-cycle reads.

\textbf{Keccak Mapping:} When executing Keccak-1600, the same physical array is reconfigured into 64x25 CBs to accommodate the 1600-bit state. Each of the 25 lanes (64-bit each) maps to columns within the CB. The $\rho$ (rotation) and $\pi$ (permutation) steps leverage implicit shifting through strategic data layout, avoiding explicit shift operations. The $\chi$ step utilizes parallel AND/XOR operations across lanes.

\textbf{NTT Mapping:} For NTT operations, the physical array adopts different logical CB configurations based on polynomial size. NTT-128 uses 128x16 CBs, while NTT-256 and larger operations utilize 256x16 CBs. A single 256x16 CB can process up to 256 coefficients. For larger NTT operations (e.g., NTT-512, NTT-1024), multiple tiles within the same row work collaboratively—the polynomial coefficients are distributed across tiles that share wordlines, enabling parallel butterfly operations across the extended data width. Twiddle factors are pre-loaded in reserved rows, accessed via bit extension commands. The radix-2 butterflies execute in $\log_2(n)$ stages with decreasing parallelism per stage.

\textbf{Kyber/Dilithium Mapping:} Both Kyber512 and Dilithium2 utilize the 256x16 CB configuration, matching NTT-256 since their polynomial operations are based on 256-element NTT. This unified configuration simplifies control logic as no reconfiguration is needed between different polynomial operations. For Keccak operations within these algorithms, the array temporarily reconfigures to 64x25 CBs. The consistent CB sizing across lattice-based algorithms enhances hardware efficiency.}

\subsection{Crypto-Near-Cache Control Commands} \label{cnc-ctrl-cmd}
The CNC is controlled by a control module that outputs a 16-bit command each cycle, which is then decoded by the CNC's control logic to manage various operations. Fig. \ref{fig:cnc-arch-mapping}(e) shows the formats of 6 command categories, which are composed of opcode, address or bit count information, and variants within the command type. Commands include writing data, logical operations (XOR/AND/OR), bitline computing (requiring an activation command before a logic operation), and bit shifts (with the middle eight bits representing the number of shifts). The \textit{ext\_bit} command uses the middle eight bits to represent the column index and a control field to specify the computing block width. 
\hlc[white]{All the control commands are pre-computed and stored in memory. When needed, they are fetched from memory to cache by control commands.}

\subsection{Error Detection and Correction} 
Though the CNC has a smaller capacity and a low soft error rate (0.7 to 7 errors per year \cite{wilkeningCalculatingArchitecturalVulnerability2014}), ECC is crucial for applications that require high reliability. 
ECC schemes are typically employed to detect and correct data written to or read from the cache. 
For logic operations in the CNC, ECC detection can check the integrity of the results by calculating the ECC of the logic result and comparing it with the corresponding ECCs of the operands \cite{agaComputeCaches2017c}. 
For shift operations, the ECC logic unit generates and stores the ECC of the shifted result for subsequent integrity checks. Cache scrubbing techniques can further help to minimize the overhead of ECC \cite{sartorCooperativeCacheScrubbing2014}.

\section{Algorithm-Architecture Co-Designing} \label{co-design}

In this section, we analyze cryptographic workloads to identify the most time-consuming operations, and \hlc[white]{list several co-designing optimization techniques that are compatible with the in-SRAM computing}, such as near-zero-cost shifting, Galois Field conversions to eliminate high-cost Look-Up Tables (LUTs), bit-parallel modular multiplication to prevent lengthy carry propagation, and hardware-supported bit extension techniques
, to maximally utilize CNC design for performance and energy-efficiency.

\subsection{Near-Zero-Cost Shifting} \label{how}

Based on our analysis in Fig. \ref{feasability}(b), Keccak computation is the most time-consuming kernel in post-quantum cryptographic algorithms, and the shift operations account for 77\% of operations.
Keccak features two types of shifts: inter-lane and intra-lane shifts, where the latter accounts for 94\% of the total shifts. 
Additionally, other kernels (e.g., AES \cite{zhangSealerInSRAMAES2022}, NTT \cite{zhangBPNTTFastCompact2023}, modular multiplication) involve numerous shift operations (on average 50\%) for operand alignment. 
For CPUs, shifting operations are conducted using separate registers and dedicated shifters. In contrast, for in-SRAM computing, shifting necessitates a bit-by-bit process which, unfortunately, compromises performance. As a result, our strategy aims to minimize the necessity for shifting as much as possible.

We propose \hlc[white]{a} \textit{implicit shifting} technique, which offers near-zero-cost shifting for the majority of the shift operations, thus avoiding large area overhead of dedicated shifters and the subsequent under-/over-utilization issues stemming from varying shifter width requirements across different cryptographic algorithms. 
This technique utilizes algorithm-specific data layout to eliminate explicit shift operations in in-memory cryptographic accelerator designs, such as inter-lane computations in Keccak and alignment among different coefficients in NTT, and can be leveraged in other applications with extensive shift operations (e.g., digital signal processing).

\subsection{Galois Field Conversion}

The S-box is a critical component of the AES algorithm. Traditional processing-in-memory designs implement the S-box using Look-Up Tables (LUTs) and substitute bytes by querying the LUT \cite{xieSecuringEmergingNonvolatile2018a}. However, incorporating LUTs leads to additional area overhead and hardware underutilization when not executing AES operations. 
To maintain the generality and flexibility of the CNC, we employ Galois Field conversion techniques to avoid introducing LUTs \cite{boyarNewCombinationalLogic2010a,yiqunzhangCompact446Gbps2016c}, which reduces large hardware overhead introduced by LUTs.
This method efficiently achieves nonlinear functions by converting computation from GF(2$ ^8$) into GF(2$ ^4$)$ ^2$. Moreover, omitting LUTs helps to reduce control overhead.

\subsection{Bit-Parallel Modular Multiplication}

Lattice-based post-quantum cryptographic algorithms extensively employ modular multiplication; however, the division involved is computationally expensive. To circumvent division, Montgomery modular multiplication \cite{montgomeryModularMultiplicationTrial1985a} is commonly used. Although this method avoids division, the high cost of multiple multiplications presents challenges for processing-in-memory designs.
To address this issue, we employ a bit-parallel Montgomery modular multiplication algorithm and extend its applicability to general scenarios, rather than being limited to cases where one of the multiplicands is known \cite{zhangBPNTTFastCompact2023}. By separately storing and computing \textit{Carry} and \textit{Sum}, and performing standard addition only at the end, this algorithm mitigates the problem of lengthy carry propagations. This design effectively leverages the parallelism offered by in-SRAM computing, thus, enhancing performance.

\section{Evaluation}\label{sec:evaluation}

In this section, we compare the proposed \textit{Crypto-Near-Cache} solution to the processing-in-memory and ASIC designs in terms of both functional properties (such as energy efficiency, performance, and area) and non-functional properties (such as generality and flexibility).

\subsection{\revisedrthree{Evaluation Methodology}}\label{sec:eval-methodology}

\revisedrthree{\textbf{Platform Selection and Comparison Rationale:} Our evaluation compares CNC (designed on RISC-V) against x86-64 CPU baselines. While this cross-architecture comparison may appear unconventional, it is justified by several factors. First, the CNC system represents a CPU baseline equipped with additional in-SRAM computing arrays and microarchitectural enhancements—essentially the same system architecture with CNC capabilities added. Although our core and cache datapath modifications are implemented on RISC-V due to its open-source nature, the fundamental design principles (distributed cache slices, virtual addressing, NoC interconnection) are readily transferable to x86 architectures. Second, x86 platforms provide mature, well-established benchmarks for cryptographic workloads, whereas RISC-V lacks comparable standardized test suites. Finally, the instruction overhead differences between ISAs are negligible compared to the computational costs of cryptographic algorithms, making platform differences minimal for our evaluation purposes.} \revisionnoter{3}{Added clear rationale for platform selection addressing R-3}

\revisedrthree{\textbf{Baseline Configuration:} Table~\ref{baseline-spec} provides detailed specifications for all evaluation platforms.} \revisionnoter{3}{Added baseline specifications table}

\added{
\begin{table}[h]
\centering
\caption{\revisedrthree{Baseline System Specifications}}
\scalebox{0.85}{
\begin{tabular}{|l|l|l|}
\hline
\textbf{Component} & \textbf{CPU Baseline} & \textbf{CNC System} \\
\hline
Processor & Intel i7-10700F (Comet Lake) & Same CPU + CNC \\
Cores/Threads & 8 cores / 16 threads & 8 cores / 16 threads \\
Max Turbo Freq. & 4.80 GHz & 4.80 GHz \\
Base Frequency & 2.90 GHz & 2.90 GHz \\
L1 Cache & 64KB per core & 64KB per core \\
L2 Cache & 256KB per core & 256KB per core \\
L3 Cache & 16MB shared & 16MB with 16 CNC arrays \\
SIMD Extensions & AVX2, AES-NI & AVX2, AES-NI + CNC ISE \\
Process Node & 14nm (normalized to 45nm) & 14nm (normalized to 45nm) \\
TDP & 65W & 65W + CNC power \\
\hline
\end{tabular}
}
\label{baseline-spec}
\end{table}
}

\textbf{General CPU Evaluation:} The evaluation uses an 8-core Intel i7-10700F with 16MB L3 cache, capable of 4.80 GHz turbo frequency. Performance is assessed by running cryptographic workloads with varying parallelism levels. Labels such as ``CPU-128'' indicate 128 concurrent encryption/decryption tasks, where each task processes independent data blocks. All algorithms use optimized implementations: AVX2 vectorization for NTT and AES-NI for AES encryption \mbox{\cite{Avanzi2017-ge,Lyubashevsky2020-cc}}. Performance metrics are captured using \textit{perf} (cycle counts) and \textit{turbostat} (dynamic power). \revisedrthree{Each experiment runs 20 iterations with 95\% confidence intervals reported. Power measurements exclude idle power and report dynamic power only. Energy efficiency is calculated as $E_{eff} = \frac{Throughput}{Power_{dynamic}}$. Statistical significance is verified using Student's t-test with p < 0.05. All results are normalized to 45nm technology for fair comparison.}\revisionnoter{3}{Provided details of statistical validation}

\textbf{CNC Evaluation:} The CNC system represents the same CPU baseline augmented with 16 CNC arrays integrated within the 16MB L3 cache (one per MB). Our evaluation methodology distinguishes between synthesized hardware metrics and simulated performance results:

\revisedrthree{\textit{Synthesis Results:} Physical design metrics (area, power, frequency) are obtained from RTL synthesis using Synopsys Design Compiler and place-and-route with Cadence Innovus at 45nm. The CNC arrays achieve 1.9GHz operation frequency. Circuit layouts use OpenRAM \cite{guthausOpenRAMOpensourceMemory2016b} for SRAM generation and PyMTL3 \cite{jiangPyMTL3PythonFramework2020a} for RTL generation.} \revisionnoter{3}{Clarified synthesis vs. simulation}

\revisedrone{\textit{Simulation Results:} Cycle counts and execution traces are obtained from our custom cycle-accurate simulator. Due to CNC's unique features (in-SRAM computing, custom instructions, cache slice integration), we developed this simulator to model CNC-specific microarchitecture including bitline operations, command execution, and cache interactions. Hardware metrics from synthesis are integrated into the simulator, and timing/power characteristics are validated against SPICE simulations for critical paths. The simulator calculates accurate cycle counts for CNC algorithm acceleration.} \revisionnoter{1}{Merged custom simulator details into simulation results subsection}

\revisedrone{\textit{Hardware Extensions:} CNC-SA configurations include dedicated shifter (64-bit barrel shifter per 64 bitlines) and adder (16-bit full adder per 16 bitlines) units. CNC-shifter and CNC-adder variants include only the respective extension. Each operation completes in a single cycle \cite{Pillmeier2002-kv,Brent1982-be}.}\revisionnoter{1}{Unified three separate CNC-SA descriptions into one location}

\textbf{\revisedrthree{Workload Composition:}} \revisionnoter{3}{Clarified workload composition details} 
\revisedrthree{Our evaluation focuses on Kyber and Dilithium algorithms with 16-bit Computing Block (CB) width. Each 512-bit wide CNC array can execute 32 parallel instances (512/16=32). With 16 CNC arrays in our baseline system, we support up to 512 parallel executions (16x32=512). The suffix numbers indicate parallelism levels: CNC-128 uses 4 arrays (128/32), CNC-256 uses 8 arrays, and CNC-512 utilizes all 16 arrays. For scalability analysis, CNC-2048 assumes 64 CNC arrays (64x32=2048), representing future large-cache systems like AMD Genoa-X with 1.1GB L3 cache. These configurations represent typical cryptographic server workloads processing multiple client requests concurrently.}

\subsection{Generality, Integration, and Flexibility}

\begin{table*}
	\centering 
	\small
	\caption{Comparison with state-of-the-art solutions w.r.t. generality, flexibility, and integration. }
	\scalebox{0.7}{%
	\begin{tabular}{>{\centering\arraybackslash}p{1.3in}>{\centering\arraybackslash}p{0.7in}>{\centering\arraybackslash}p{0.3in}>{\centering\arraybackslash}p{1.6in}>{\centering\arraybackslash}p{0.4in}>{\centering\arraybackslash}p{0.8in}>{\centering\arraybackslash}p{0.65in}>{\centering\arraybackslash}p{0.8in}}
		\toprule \midrule
		& Design & Generality & Evaluated Kernels$^1$ & Integration & Max-Bitwidth$^2$ & Max Order$^3$ & TEE Expansion  \\ \hline \hline 
  \addlinespace
        Crypto-Near-Cache & near-Cache & \multirow{2}{*}{\centering \tikz\draw[OliveGreen,fill=OliveGreen] (0,0) circle (1ex);} & AES, SHA3, NTT; & \multirow{2}{*}{\centering \tikz\draw[OliveGreen,fill=OliveGreen] (0,0) circle (1ex);} & \multirow{2}{*}{512} & \multirow{2}{*}{9k} & \multirow{2}{*}{\xmark} \\
        (CNC)  & in-SRAM & &  Kyber, Dilithium & & & & \\ \addlinespace
        \hline
        \addlinespace
        ReCryptor \cite{zhangRecryptorReconfigurableCryptographic2018a} & in-Cache & \tikz\draw[OliveGreen,fill=OliveGreen] (0,0) circle (1ex);  & AES, SHA3  &  \tikz\draw[BrickRed,fill=BrickRed] (0,0) circle (1ex); & N/A & N/A & \xmark \\
        MeNTT \cite{liMeNTTCompactEfficient2022e} & in-SRAM & \tikz\draw[OliveGreen,fill=OliveGreen] (0,0) circle (1ex); & NTT  &  \tikz\draw[BrickRed,fill=BrickRed] (0,0) circle (1ex);  & 32 & 32k & \cmark \\
        IMCRYPTO \cite{reisIMCRYPTOInMemoryComputing2022a} & near-Cache  & \tikz\draw[YellowOrange,fill=YellowOrange] (0,0) circle (1ex); & AES, SHA-2  &  \tikz\draw[OliveGreen,fill=OliveGreen] (0,0) circle (1ex);  & N/A & N/A & \xmark\\
        AIM \cite{xieSecuringEmergingNonvolatile2018a} & in-NVM & \tikz\draw[BrickRed,fill=BrickRed] (0,0) circle (1ex); & AES &  \tikz\draw[BrickRed,fill=BrickRed] (0,0) circle (1ex); & N/A & N/A & \cmark   \\
        FeCrypto \cite{liuFeCryptoInstructionSet2022} & in-FeFET & \tikz\draw[YellowOrange,fill=YellowOrange] (0,0) circle (1ex);  & AES, SHA-1, MD5, SM3 &  \tikz\draw[YellowOrange,fill=YellowOrange] (0,0) circle (1ex); & N/A & N/A & \cmark \\
        CryptoPIM \cite{nejatollahiCryptoPIMInmemoryAcceleration2020a} & in-ReRAM & \tikz\draw[BrickRed,fill=BrickRed] (0,0) circle (1ex); & NTT &  \tikz\draw[BrickRed,fill=BrickRed] (0,0) circle (1ex); & 32 & 32k & \cmark   \\ 
        RM-NTT \cite{parkRMNTTRRAMBasedComputeinMemory2022b} & in-ReRAM & \tikz\draw[BrickRed,fill=BrickRed] (0,0) circle (1ex); & NTT &  \tikz\draw[OliveGreen,fill=OliveGreen] (0,0) circle (1ex);  & 16 & 1k & \cmark  \\ \addlinespace
        \hline
        \addlinespace
        RePQC \cite{chenReportPostQuantumCryptography2016a} & ASIC & \tikz\draw[BrickRed,fill=BrickRed] (0,0) circle (1ex); & SHA3, NTT; Kyber, Dilithium &  \tikz\draw[OliveGreen,fill=OliveGreen] (0,0) circle (1ex);  & 24 & 256 & \cmark  \\
        ISSCC'15 \cite{verbauwhede24CircuitChallenges2015a} & ASIC & \tikz\draw[BrickRed,fill=BrickRed] (0,0) circle (1ex); & NTT &  \tikz\draw[OliveGreen,fill=OliveGreen] (0,0) circle (1ex); & 14 & 256 & \cmark   \\
        LEIA \cite{songLEIA05mm2140mW2018a} & ASIC & \tikz\draw[BrickRed,fill=BrickRed] (0,0) circle (1ex); & SHA-2, NTT &  \tikz\draw[OliveGreen,fill=OliveGreen] (0,0) circle (1ex); & 32 & 2k & \cmark   \\
        Sapphire \cite{banerjeeSapphireConfigurableCryptoProcessor2019b} & ASIC  & \tikz\draw[BrickRed,fill=BrickRed] (0,0) circle (1ex); & SHA3, NTT; Kyber, Dilithium &  \tikz\draw[OliveGreen,fill=OliveGreen] (0,0) circle (1ex);  & 24 & 1k & \cmark \\
        VPQC \cite{xinVPQCDomainSpecificVector2020} & ASIC & \tikz\draw[BrickRed,fill=BrickRed] (0,0) circle (1ex); & SHA3, NTT; Kyber &  \tikz\draw[OliveGreen,fill=OliveGreen] (0,0) circle (1ex); & 16 & 2k & \cmark   \\
        \addlinespace
	 \hline

		\bottomrule
		\addlinespace
	\end{tabular}
	}%
	\label{overalltable}
\raggedright\small \\ \emph{1.} Set of kernels and NIST-selected algorithms that are evaluated in each study. \\
\emph{2.} Maximum coefficient bitwidth that the design support for NTT computation. \\
\emph{3.} Maximum polynomial order that the design can support for NTT computation. 
\end{table*}

\hlc[white]{As shown in Table \mbox{\ref{overalltable}}}, In-SRAM solutions like CNC, ReCryptor \cite{zhangRecryptorReconfigurableCryptographic2018a}, and MeNTT \cite{liMeNTTCompactEfficient2022e}, use bitwise logical operations, making them adaptable for various algorithms. Contrastingly, Near-Cache and ASIC designs with dedicated units may face reconfiguration challenges and could be less general. 
Though solutions like RePQC \cite{chenReportPostQuantumCryptography2016a} and Sapphire \cite{banerjeeSapphireConfigurableCryptoProcessor2019b} support diverse post-quantum cryptography kernels, they may not be as adaptable to other or future cryptographic algorithms \cite{ghoshLightweightPostQuantumSecureDigital2019,kumarSecuringFutureInternet2022b,mustafaLightweightPostQuantumLatticeBased2020}. Despite its generic computing units, the Near-cache design like IMCRYPTO \cite{reisIMCRYPTOInMemoryComputing2022a} impose significant control costs on the CPU (Section \ref{imcrypto}).

When considering integration, solutions like RM-NTT \cite{parkRMNTTRRAMBasedComputeinMemory2022b} and ASIC-based designs can be effortlessly connected to existing systems. However, In-memory designs require extensive data preprocessing, disrupting seamless integration. Near-cache designs like CNC and IMCRYPTO effectively use existing data access mechanisms, allowing efficient acceleration of applications with minor rearrangements.

Flexibility in cryptographic hardware accelerator designs is essential. CNC stands out in this aspect by supporting various configurations of each algorithm, accommodating up to 512-bit wide and 9k-order polynomial NTTs. In general, in-memory designs tend to offer more flexibility than ASIC designs.

\textit{In summary, CNC emerges as an optimal choice based on its exceptional generality, ease-of-integration, and flexibility. It supports various algorithms and settings, and can seamlessly integrate into current systems without modifying existing memory access mechanisms. As an additional benefit, being an on-chip solution, CNC keeps all computations within the processor boundary.}

\subsection{Energy Efficiency and Performance Comparison}\label{sec:eval-comparison}
\begin{figure}[htp]
	\centerline{\includegraphics[width=5.4in]{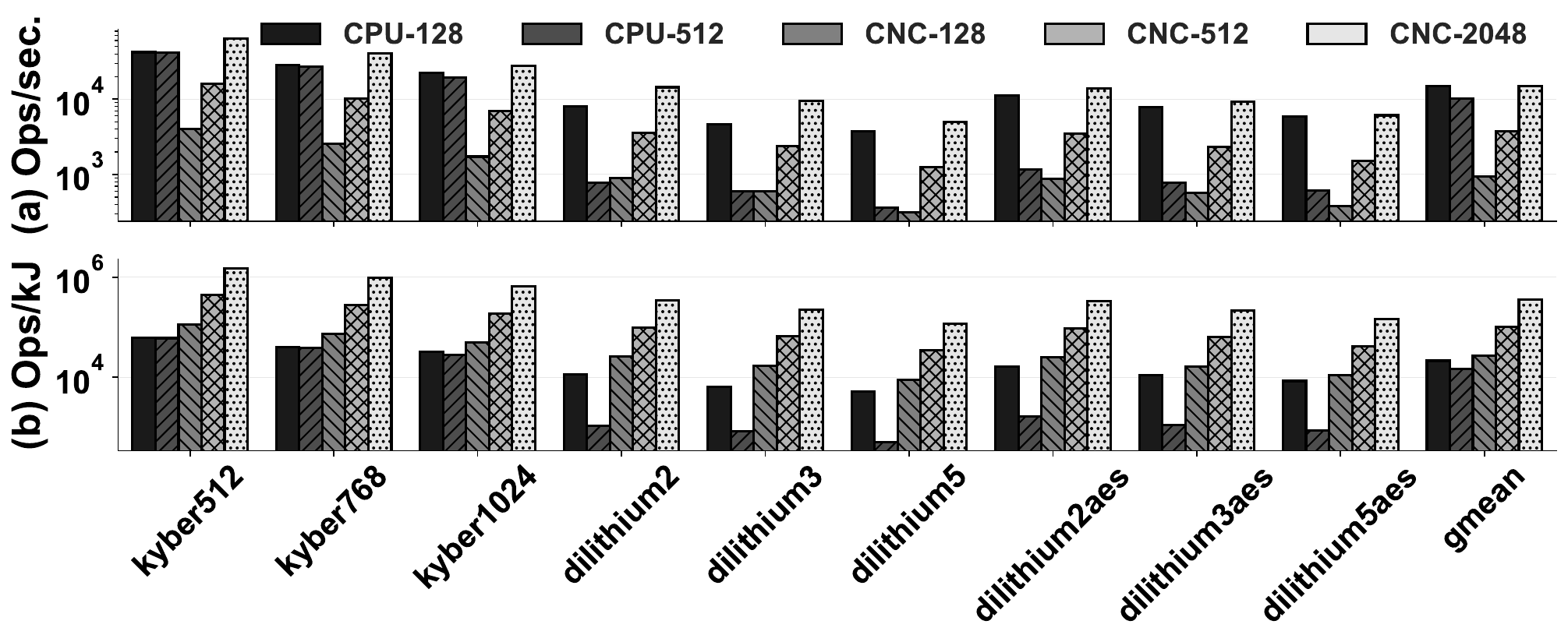}}
	\caption{\revisedrone{Performance and energy efficiency comparison of CNC against CPU baselines across PQC algorithms: (a) throughput in operations per second, and (b) energy efficiency in operations per kilojoule.}\revisionnoter{1}{Split original figure into separate figure (CNC vs CPU) and table (CPU vs CNC vs CNC-SA) for clarity}}
	\label{fig:performance-comparison}
\end{figure}

\revised{Fig. \ref{fig:performance-comparison} presents comprehensive performance and energy efficiency comparisons between CNC and CPU implementations across PQC algorithms. The suffix numbers denote parallelism levels (e.g., CNC-128 indicates 128 parallel cryptographic operations).

For throughput (Fig. \ref{fig:performance-comparison}(a)), CNC-2048 achieves superior performance compared to both CPU-128 and CPU-512, demonstrating approximately 1.5$\times$ higher throughput than CPU-128 for Kyber512 and up to 1.8$\times$ for Dilithium2. While CPU implementations show competitive throughput at lower parallelism levels, this comes at the cost of significantly lower energy efficiency.

For energy efficiency (Fig. \ref{fig:performance-comparison}(b)), CNC architectures exhibit substantial improvements, with CNC-2048 achieving up to 25$\times$ better energy efficiency compared to CPU-128 for Kyber512 and 30$\times$ for Dilithium2. This performance boost stems from: (1) combined data storage and computation within SRAM arrays, reducing data movement energy; (2) stall-free execution with fixed order, eliminating complex instruction scheduling and branch prediction; (3) efficient scaling where energy efficiency is maintained as parallelism increases, contrasting with CPU architectures where efficiency decreases due to heightened scheduling demands; and (4) minimal CPU involvement—while algorithms like Kyber and Dilithium require CPU for rejection sampling, energy analysis shows CPU consumes only 0.12\% of total energy in CNC. The independent operation of each CNC slice enables near-linear throughput scaling without proportional energy cost increases.} \revisionnoter{1}{Merged throughput and energy efficiency discussions}

\begin{table}[htp]
\centering
\caption{\revisedrone{Performance and energy efficiency comparison across CPU, CNC, and CNC-SA configurations.}\revisionnoter{1}{Created from split to provide detailed comparison including CNC-SA hardware extensions}}
\scalebox{0.82}{
\begin{tabular}{|l|c|c|c|c|c|}
\hline
\multirow{2}{*}{\textbf{Configuration}} & \multirow{2}{*}{\textbf{Parallelism}} & \multicolumn{2}{c|}{\textbf{Throughput (Ops/s)}} & \multicolumn{2}{c|}{\textbf{Energy Efficiency (Ops/kJ)}} \\
\cline{3-6}
 & & \textbf{Kyber512} & \textbf{Dilithium2} & \textbf{Kyber512} & \textbf{Dilithium2} \\
\hline
CPU-128 & 128 & 42,100 & 8,050 & 60,700 & 11,300 \\
CPU-512 & 512 & 41,900 & 772 & 60,300 & 1,080 \\
\hline
CNC-128 & 128 & 4,020 & 901 & 114,723 & 25,721 \\
CNC-512 & 512 & 16,081 & 3,605 & 440,259 & 98,706 \\
CNC-2048 & 2048 & 64,324 & 14,421 & 1,514,988 & 339,659 \\
\hline
CNC-128-SA$^*$ & 128 & 5,434 & 1,383 & 155,063 & 39,473 \\
CNC-512-SA$^*$ & 512 & 21,736 & 5,533 & 595,070 & 151,479 \\
CNC-2048-SA$^*$ & 2048 & 86,943 & 22,132 & 2,047,711 & 521,260 \\
\hline
\end{tabular}
}
\begin{center}
\parbox{0.85\linewidth}{
\small $^*$CNC-SA configurations include hardware extensions as described in the evaluation methodology section.
}
\end{center}
\label{tab:cnc-comparison}
\end{table}

Table \ref{tab:cnc-comparison} presents a comprehensive comparison of CPU, CNC, and CNC-SA configurations. The CNC-SA variant with hardware extensions achieves approximately 1.35$\times$ speedup for Kyber512 and 1.54$\times$ for Dilithium2 over base CNC, executing shift and arithmetic operations in single cycles. Energy efficiency improvements are substantial: CNC-2048-SA achieves 2.05M operations per kilojoule for Kyber512 (34$\times$ over CPU-128) and 521k operations per kilojoule for Dilithium2 (46$\times$ over CPU-128).

\subsection{Latency Breakdown} \label{latency-bd}

\begin{figure}[tp]
\centerline{\includegraphics[width=5.4in]{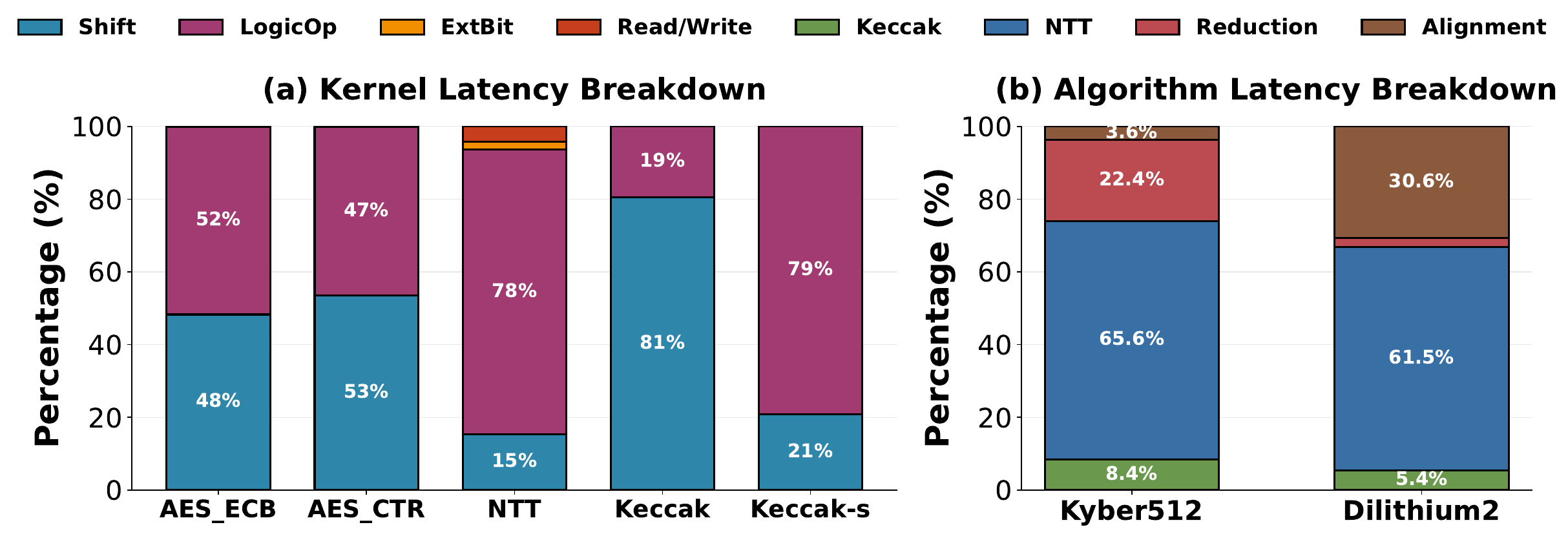}}
\caption{\revisedrone{Latency breakdown analysis: (a) Different cryptographic kernels on CNC showing the distribution of shift operations, logic operations, extended bit operations, and read/write operations. (b) Kyber512 and Dilithium2 algorithms showing the contribution of different computational components.}\revisionnoter{1}{Changed from pie charts to stacked bar charts for clearer visualization of operation distributions}}
\label{fig:latency-breakdown}
\end{figure}

We first conduct a breakdown analysis of kernel latency, as illustrated in Fig. \ref{fig:latency-breakdown}(a). We observe that the number of logical and shift operations in AES is nearly equal due to the utilization of Galois Field conversion, which allows logical operations to replace LUT queries. In NTT, shift operations account for only 15\% of the total cycles, stemming from the employment of bit-parallel modular multiplication to avoid carry propagation. For Keccak, approximately 80\% of operations are shifts, primarily intra-lane shifts, as inter-lane shifts are mitigated by \textit{implicit shifting} (resulting in a 94\% shift reduction). The high proportion of shifts in Keccak can be attributed to the significant reduction of logic operations. With the use of large vector computing units in CNC, the number of logical operations is considerably reduced, which leads to a 95\% reduction in total latency compared to traditional in-cache designs \cite{agaComputeCaches2017c}. To further optimize Keccak, incorporating a dedicated rotate shifter (\textit{Keccak-s} in Fig. \ref{fig:latency-breakdown}(a)) could potentially reduce the shift portion to 6\%.

As shown in Fig. \ref{fig:latency-breakdown}(b), besides NTT, the largest share Kyber cycles is attributed to \textit{Reduction} kernel due to the additional reduction operation in inverse-NTT.
Dilithium has lower Reduction overhead but a larger signature size, necessitating more operations for packing and range converting signatures (categorized as \textit{Alignment}).
Users with specific requirements can add dedicated modules (e.g., Barrett reduction or packing module) to further accelerate their target applications, as demonstrated in Section~\ref{sec:ablation}.

\subsection{\hlc[white]{Ablation Analysis}} \label{sec:ablation}

\hlc[white]{In our ablation study, we analyzed the impact of co-design algorithms and dedicated peripheral circuits on performance across various cryptographic primitives.}

\hlc[white]{\textbf{AES:}
In AES, the performance degradation with the Zero Cost Shift (ZCS) technique becomes evident. While wider vectors expedite the \mbox{\textit{AddRoundKey}} phase, they lead to a performance drop when operations occur between bytes due to the overhead from shift operations. Introducing a LUT can enhance CNC's performance in AES applications, as the \textit{SubByte} operation can be completed by accessing the Sbox in the LUT alone. Yet, due to its significant hardware overhead and challenges in supporting multiple parallel computation processes, LUT wasn't integrated into the CNC. Further inclusion of a dedicated shifter to the CNC shows significant performance improvement in AES, given the pervasiveness of shift operations across its various stages.}

\hlc[white]{\textbf{Keccak:}
For Keccak, the lack of ZCS necessitates an adjustment in data layout when entering the Keccak round, incurring additional overhead. However, a remarkable performance enhancement is observed when a dedicated shifter is incorporated on the periphery of CNC. This is due to Keccak's extensive intra-lane shift operations, which the dedicated shifter can execute efficiently.}

\hlc[white]{\textbf{NTT:}
Regarding NTT, the exclusion of Hardware-Supported Bit Extension (EBCC) mandates bit-by-bit shifting for sign bit propagation, causing a performance drop. The performance further deteriorates when both Bit-Parallel Modular Multiplication and Zero Cost Shift (BPMM \& ZCS) are omitted, as calculations become bit-serial, amplifying the latency. On the other hand, adding a dedicated shifter to CNC doesn't benefit NTT's performance, as all its shift operations happen between adjacent bitlines within a single cycle. However, the integration of a full adder significantly boosts performance, given its ability to handle bit-by-bit carry propagation within one cycle when performing sign conversion.}

\begin{figure}[htp]
\centerline{\includegraphics[width=3.8in]{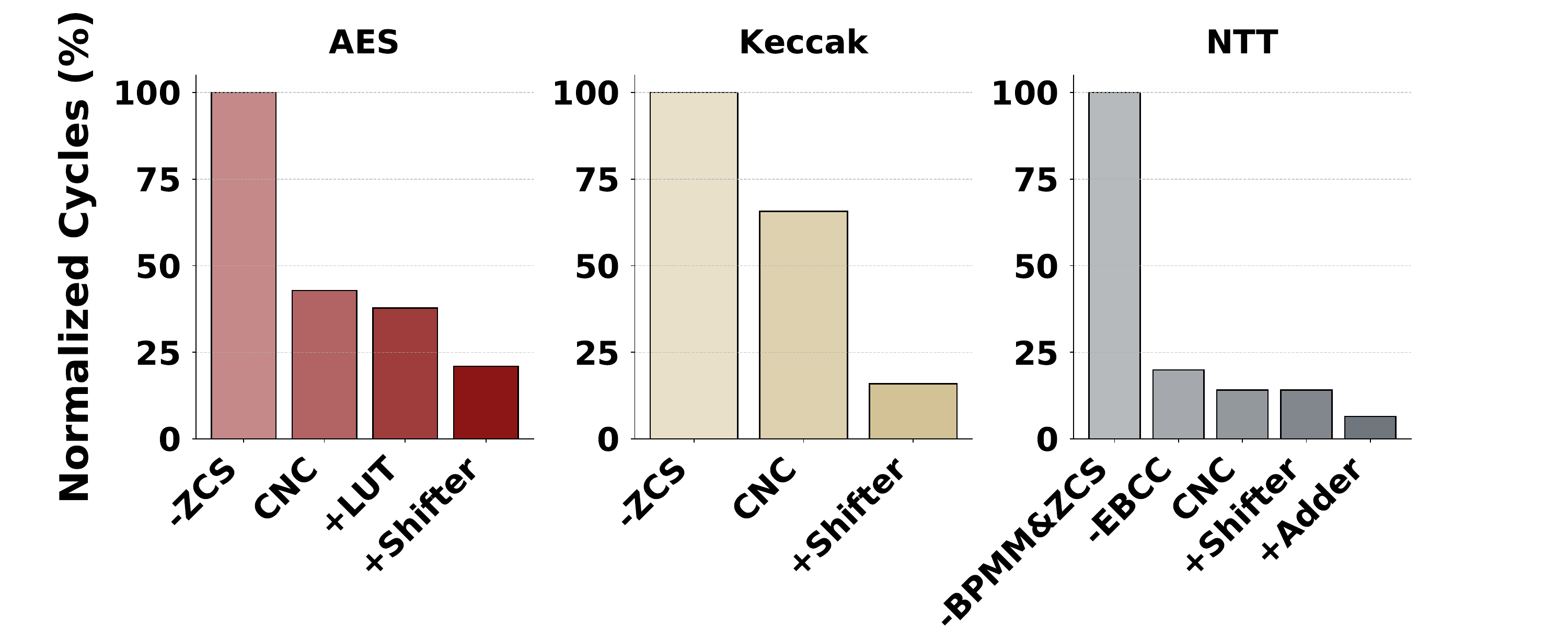}}
\caption{\hlc[white]{Analysis of co-design algorithms and dedicated peripheral logics over CNC.}
}
\label{ablation}
\end{figure}

\subsection{Data Movement Analysis}

CNC reduces data movement by reusing commands for building block kernels such as AES, modular multiplication, and Keccak. Data movement only occurs when initially fetching inputs and writing results to the data cache. Table \ref{command-overhead} shows the storage capacity required for different kernels, with the most demanding Keccak round requiring 13.6kB storage capacity, which can be satisfied by a command array of size 256$\times$512. The table also provides a detailed breakdown of control overhead for various cryptographic functions. For AES-128, the total command count is 5,900 instructions requiring 11.5kB of storage. The most instruction-intensive operations are ShiftRows (456 instructions) and SubBytes (357 instructions). By reusing kernel commands stored in the command array, data movement is reduced, especially for common cryptographic workloads that require consecutive Keccak (reduced by 11$\times$) and NTT operations (reduced by 611$\times$).

\begin{table}[tp]
	\centering
	\caption{\revisedrtwo{Command storage requirements for CNC kernels.}\revisionnoter{2}{Added detailed breakdown of command requirements for individual cryptographic functions to clearly demonstrate our implementation}}
	\scalebox{0.85}{
	\begin{tabular}{|l|c|c|c|}
		\hline
		\textbf{Kernel/Function} & \textbf{\#Commands} & \textbf{Capacity (KB)} & \textbf{Iterations} \\
		\hline
		\multicolumn{4}{|c|}{\textit{Summary Statistics}} \\
		\hline
		AES-128 (total) & 5,900 & 11.5 & - \\
		16-bit MMult & 900 & 1.9 & - \\
		32-bit MMult & 1,900 & 3.7 & - \\
		Keccak & 7,000 & 13.6 & - \\
		\hline
		\multicolumn{4}{|c|}{\textit{\revisedrtwo{Detailed Breakdown}}} \\
		\hline
		AES: BitSlicing & 288 & 0.58 & 2 \\
		AES: AddRoundKey & 24 & 0.05 & 11 \\
		AES: SubBytes & 357 & 0.71 & 10 \\
		AES: ShiftRows & 456 & 0.91 & 10 \\
		AES: MixColumns & 258 & 0.52 & 9 \\
		\hline
		GHASH: ByteArrange & 63 & 0.13 & 1 \\
		GHASH: ByteAligning & 138 & 0.28 & 8 \\
		GHASH: GaloisMult & 16 & 0.03 & 1,024 \\
		\hline
		SHA3: StatePermute & 633 & 1.27 & 24 \\
		\hline
		\textbf{Total (Detail)} & 2,233 & 4.47 & - \\
		\hline
	\end{tabular}}
	\label{command-overhead}
\end{table}

\subsection{Comparison with In-Cache and ASIC Solutions } \label{compare-pim-asic}

This section compares CNC's performance and energy efficiency with in-cache and ASIC designs for various kernels. Table \ref{kernel-comparison} shows a comprehensive comparison across Keccak and NTT kernels. For Keccak, though Inhale surpasses CNC-shifter in frequency and energy efficiency due to its smaller array size and dedicated rotate shifters, CNC-shifter outperforms in throughput. By adding rotate shifter units to CNC, it achieves 5.4$\times$ higher throughput than Inhale, with similar energy efficiency. For NTT operations, CNC-adder shows higher throughput than BP-NTT \cite{zhangBPNTTFastCompact2023} and competitive performance against ASIC designs, despite being a more general framework. CNC is proposed as a general near-cache framework that can be enhanced with dedicated peripheral logic for specific applications (Section~\ref{sec:ablation}).

\begin{table}[htp]
	\centering
	\caption{\revisedadd{Performance comparison of CNC with state-of-the-art kernel-specific designs.} \revisionnoteadd{Added detailed comparison table to clarify different near-cache design approaches and justify our slice-level choice}}
	\scalebox{0.82}{
	\begin{tabular}{|l|l|c|c|c|c|}
		\hline
		\textbf{Kernel} & \textbf{Design} & \textbf{Freq. (GHz)} & \textbf{Area (mm$^2$)} & \textbf{Throughput} & \textbf{Efficiency} \\
		\hline
		\multirow{2}{*}{Keccak} & CNC-shifter & 1.9 & 3.1 & 4.3 MOps & 34.8 kOp/mJ \\
		& Inhale \cite{zhangInhaleEnablingHighPerformance2022b} & 2.6 & 0.4 & 0.8 MOps & 37.4 kOp/mJ \\
		\hline
		\multirow{6}{*}{NTT} & CNC-adder & 1.9 & 3.4 & 1.5 kOps & 12.2 kOp/mJ \\
		& BP-NTT \cite{zhangBPNTTFastCompact2023} & 2.6 & 0.35 & 178 Ops & 8.6 kOp/mJ \\
		& Sapphire \cite{banerjeeSapphireConfigurableCryptoProcessor2019b} & 0.06 & 0.35 & 49.7 Ops & 4.2 kOp/mJ \\
		& LEIA \cite{songLEIA05mm2140mW2018a} & 0.27 & 1.8 & 1.7 kOps & 22.7 kOp/mJ \\
		& RePQC \cite{chenReportPostQuantumCryptography2016a} & 0.31 & 9.3 & 9.7 kOps & 10.2 kOp/mJ \\
		& VPQC \cite{xinVPQCDomainSpecificVector2020} & 0.19 & 3.1 & 4.1 kOps & 9.8 kOp/mJ \\
		\hline
	\end{tabular}}
	\begin{center}
	\parbox{0.85\linewidth}{
	\small MOps: Million operations per cycle; kOps: Thousand operations per cycle; kOp/mJ: Thousand operations per millijoule.
	}
	\end{center}
	\label{kernel-comparison}
\end{table}

\subsection{Comparison with IMCRYPTO}\label{imcrypto}
IMCRYPTO \cite{reisIMCRYPTOInMemoryComputing2022a} locates its compute module outside the cache (Fig.~\ref{nearInCache}(c)), thus relying on limited external bandwidth for data movement. In contrast, CNC harnesses higher internal cache bandwidth for greater efficiency. Another difference lies in the ISA granularity and control flow: IMCRYPTO requires the CPU to issue each instruction to the compute-enabled array, needing 2048 instructions for SHA-256, whereas CNC stores the algorithm in a command array and needs only two instructions (one to load commands, another to initiate execution).
Our quantitative comparison reveals significant performance advantages: for AES-128 operations with identical array size (256×512) and process node (45nm), IMCRYPTO's bit-serial layout results in approximately 900µs latency compared to CNC's 20.6µs—a 44× higher latency. Additionally, CNC achieves 198.6 Mbps throughput versus IMCRYPTO's 160 Mbps, demonstrating about 20\% better throughput thanks to its bit-parallel design and co-optimized algorithms.

\subsection{Potential Limitations of Our Proposal}\label{sec:limitations}

Integrating CNC modules increases the chip area compared to designs without such enhancements. For instance, our CNC-512 occupies 24.32 mm$ ^2$, which is about 1\% of the i9-7900X processor's area. Despite this overhead, our use of bit-parallel computing reduces the required array size compared to bit-serial designs, mitigating some area impact. While CNC significantly improves performance over general-purpose processors, it may not achieve the same low latency as specialized ASIC accelerators optimized for specific algorithms; however, at the same area consumption, CNC's latency is 43.6$\times$ lower than that of IMCRYPTO. 

\revisedrfour{From a security perspective, while CNC's architectural properties include keeping all computations on-chip and minimizing data movement, we acknowledge several limitations. CNC does not implement hardware-level countermeasures against side-channel attacks such as masking or hiding techniques. The current design relies on software-based protections and constant-time algorithm implementations. Formal security analysis and comprehensive side-channel resistance evaluation remain as future work. We note that CNC's primary contribution is in performance and energy efficiency for cryptographic workloads, with security benefits being secondary to these main objectives.} \revisionnoter{4}{Toned down security claims as requested in R-4}

Additionally, integrating CNC modules requires modifications to the core and cache datapath, which entails additional testing and verification efforts \cite{Golmohamadi2020-pe, Bhosle2023-lx}.
Extending CNC to efficiently accelerate non-cryptographic workloads requires careful algorithm-hardware co-design, which is challenging and part of our future work.
\section{Conclusion}
This paper presents Near-Cache-Slice Computing, or referred to as Crypto-Near-Cache (CNC), to provide a secure, general, flexible, low-overhead, energy-efficient, and easy-to-integrate solution for accelerating pre-/post-quantum cryptography algorithms and beyond.
By carefully designing core datapath, cache datapath, and ISA extensions, coupled with optimization techniques for algorithm co-designing (e.g., near-zero-cost shifting, Galois Field conversion, bit-parallel modular multiplication, and bit extension), the proposed architecture achieves considerable energy efficiency and throughput. 





\bibliographystyle{alpha}
\bibliography{paperpile,references}

\end{document}